\documentclass[aps, twocolumn, letterpaper, superscriptaddress]{revtex4}

\usepackage{amsmath}
\usepackage{amssymb}
\usepackage{mathrsfs}
\usepackage{xspace}
\usepackage{braket}
\usepackage{bbold}
\usepackage{color}

\usepackage{ifpdf}
\ifpdf
\fi

\newcommand{\eq}[1]{(\ref{#1})}
\newcommand{\Eq}[1]{Eq.~(\ref{#1})}
\newcommand{\Eqs}[1]{Eqs.~(\ref{#1})}

\newcommand{\Sec}[1]{Sec.~\ref{#1}}
\newcommand{\Ref}[1]{Ref.~\cite{#1}}
\newcommand{\Refs}[1]{Refs.~\cite{#1}}
\newcommand{\App}[1]{Appendix~\ref{#1}}

\newcommand{\eg}{{e.g.,\/}\xspace}
\newcommand{\ie}{{i.e.,\/}\xspace}

\newcommand{\pd}{\partial}
\newcommand{\del}{\nabla}

\newcommand{\mc}[1]{\mathcal{#1}}
\newcommand{\mcc}[1]{\mathfrak{#1}}
\newcommand{\msf}[1]{\mathsf{#1}}
\newcommand{\mcu}[1]{\mathscr{#1}}
\renewcommand{\vec}[1]{{\boldsymbol{\rm #1}}}
\newcommand{\oper}[1]{\smash{\widehat{#1}}}
\newcommand{\boper}[1]{\oper{\vec{#1}}}
\newcommand{\hilb}[1]{\mc{#1}}

\newcommand{\Deff}{\msf{D}}
\newcommand{\weyl}{\mathscr{W}}
\newcommand{\field}{\Psi}
\newcommand{\envel}{\psi}
\newcommand{\Deffr}{\Deff_H}
\newcommand{\Deffi}{\Deff_A}
\newcommand{\sigmap}{\tilde{\sigma}}

\newcommand{\boldparen}[1]{\textbf{(}#1\textbf{)}}
\newcommand{\tLambda}{\Gamma}

\newcommand{\vmetr}{\gamma}
\newcommand{\mdet}[1]{#1_\diamond}
\newcommand{\dd}{\mathrm{d}}

\newcommand{\new}[1]{#1}

\begin{document}

\title{Quasioptical modeling of wave beams with and without mode conversion:\\I.~Basic theory}

\author{I.~Y. Dodin}
\affiliation{Princeton Plasma Physics Laboratory, Princeton, New Jersey 08543, USA} 
\author{D.~E. Ruiz}
\affiliation{Sandia National Laboratories, P.O. Box 5800, Albuquerque, New Mexico 87185, USA}
\author{K. Yanagihara}
\affiliation{Nagoya University, 464-8601, Nagoya, Aichi, Japan}
\author{Y. Zhou}
\affiliation{Princeton Plasma Physics Laboratory, Princeton, New Jersey 08543, USA} 
\author{S. Kubo}
\affiliation{National Institute for Fusion Science, National Institutes of Natural Sciences, 509-5292, Toki, Gifu, Japan}

\begin{abstract}
This work opens a series of papers where we develop a general quasioptical theory for mode-converting electromagnetic beams in plasma and implement it in a numerical algorithm. Here, the basic theory is introduced. We consider a general quasimonochromatic multi-component wave in a weakly inhomogeneous linear medium with no sources. For any given dispersion operator that governs the wave field, we explicitly calculate the approximate operator that governs the wave envelope $\envel$ to the second order in the geometrical-optics parameter. Then, we further simplify this envelope operator by assuming that the gradient of $\envel$ transverse to the local group velocity is much larger than the corresponding parallel gradient. This leads to a parabolic differential equation for $\envel$ (``quasioptical equation'') in the basis of the geometrical-optics polarization vectors. Scalar and mode-converting vector beams are described on the same footing. We also explain how to apply this model to electromagnetic waves in general. In the next papers of this series, we report successful quasioptical modeling of radiofrequency wave beams in magnetized plasma based on this theory.
\end{abstract}

\maketitle

\bibliographystyle{full}

\section{Introduction}
\label{sec:intro}

\new{\subsection{Motivation}}
\label{sec:motivation}

Describing the propagation of waves in inhomogeneous media is a classic problem with a long history \cite{book:tracy, book:kravtsov, book:whitham}. It is particularly important in fusion research, where quasistationary beams of electromagnetic (EM) radiation are commonly used for many purposes and need to be modeled with fidelity \cite{book:stix}. Full-wave modeling, which involves solving the complete Maxwell's equations, can be impractical at short (cm and mm) wavelengths, especially when multi-dimensional simulations with complex geometries are required, such as those of tokamak and stellarator plasmas. Hence, reduced methods have been widely used in practice. These methods are rooted in geometrical optics (GO) \cite{book:tracy} and include conventional ray tracing \new{\cite{ref:bonoli14, tex:genray, ref:marushchenko14, ref:tsujimura15}}, complex ray tracing \cite{ref:mazzucato89, ref:farina07}, beam tracing \new{\cite{ref:pereverzev92, ref:pereverzev98, ref:poli01, ref:poli18}}, and variations of thereof \cite{foot:thereof}. There are also other ``quasioptical'' models, such as in \Refs{ref:balakin07a, ref:balakin07b, foot:balakin, foot:phillips86}, that resolve the evolution of the beam transverse structure without adopting any particular ansatz for the intensity profile. Still, they assume that only one branch of the dispersion relation is excited in each given case \cite{foot:raycon}, \new{\ie mode conversion does not occur \cite{foot:mc}. As a result, today's simulations of mode-converting beams mainly rely on full-wave codes \cite{foot:phillips86} and thus have to compromise the fidelity by reducing the number of the dimensions resolved \cite{ref:tsujimura15}.}

\new{However, simulating mode conversion does not actually require the full-wave approach; in fact, it can be done within the quasioptical approach too, if the latter is generalized properly. Developing the corresponding algorithms would facilitate not only fusion applications. For example, advanced quasioptical models could find use in optics and general relativity, where mode-converting beams are also possible and have been attracting much attention lately \cite{ref:bliokh15, foot:marius}. Hence, it is potentially beneficial to approach the problem of quasioptical modeling as a general-physics problem, without restricting it to fusion applications.} 

\new{\subsection{General idea}}

\new{Mode conversion can be described quite generally, particularly without restricting it to narrow regions in space, within ``extended geometrical optics'' (XGO), which was developed recently \cite{phd:ruiz17, my:covar, my:qdirac, my:qdiel, my:qdirpond}.} XGO is a theory that calculates the leading-order correction $U$ to the GO dispersion operator of a \textit{general} vector wave and shows \cite{my:qdirac} that this correction is analogous to (and a generalization of) the Stern--Gerlach Hamiltonian of a quantum spin-$1/2$ electron. Accordingly, $U$ is responsible for two effects simultaneously: (i) it modifies the ray equations just like spin--orbital interactions affect the electron motion, and (ii) it also governs mode conversion, which appears as a direct analog of spin up--down transitions \cite{my:xo}. An examination of the quasioptical algorithms such as those in \Refs{ref:pereverzev98, ref:balakin07a} shows that they already involve calculations of terms similar to $U$. Hence, adding the mode-conversion capability to quasioptical codes should not be burdensome and is not expected to slow down the codes considerably. However, formulating the corresponding theory is easier to do using the abstract quantumlike formalism of XGO. One only needs to upgrade the existing XGO by adding diffraction, for which it also helps to introduce more general coordinates with curved metric~\cite{foot:covar}.

In this series of papers (Papers I-III), we propose such an upgrade of XGO and apply it to numerical simulations. \new{Our goal is to develop a general modular framework which later could be applied to a broad variety of problems both in plasma physics and beyond. Within this framework to be presented, one can assume general dispersion, diffraction, and polarization effects, including mode conversion of not just two but arbitrarily many resonant waves. Also importantly, our formulation below is not restricted to EM waves, since we do not specify the dispersion operator in the governing wave equation.}

\new{Our series of papers is organized as follows. In Paper~I, we introduce the basic theory of waves that diffract and mode-convert simultaneously. In Paper~II \cite{tex:mypaper2} and Paper~III \cite{tex:mypaper3}, we apply this theory to perform quasioptical modeling of radiofrequency-wave beams in magnetized plasma as an example. In particular, we consider applications to mode conversion caused by magnetic shear in edge plasma \cite{my:xo}, which, for example, is a known problem \cite{ref:notake05, ref:kubo15} in the Large Helical Device \cite{ref:iiyoshi99, ref:osakabe17}.}

\mbox{}

\subsection{Outline}

In this first paper of our series, we consider an arbitrary quasimonochromatic multi-component wave in a weakly inhomogeneous linear medium. Supposing that $\oper{D}$ is some dispersion operator governing the wave dynamics, we simplify $\oper{D}$ and obtain an approximate operator that governs the wave envelope $\envel$. Then, we derive a parabolic differential equation for $\envel$ (``quasioptical equation'') by assuming that the gradient of the wave envelope transverse to the local group velocity is much larger than the corresponding parallel gradient. The resulting theory applies to both scalar and mode-converting vector beams. At the end of the paper, we also discuss how this model can be applied to EM waves in particular. However, readers who are mainly interested in simulations as opposed to the general theory are encouraged to proceed straight to Paper~II, where our key equations are overviewed in a simplified form and without derivations.

This paper is organized as follows. In \Sec{sec:gen}, we introduce the general problem. In \Sec{sec:env}, we formalize the concept of the envelope dispersion operator. In \Sec{sec:weylexp}, we derive an approximation for the envelope dispersion operator for scalar waves, and we also derive its quasioptical approximation. In \Sec{sec:vector}, we extend this model to vector waves. In \Sec{sec:EM}, we explain how to apply the resulting theory to EM waves in particular. In \Sec{sec:conc}, we present our main conclusions. In \App{app:sumnot}, we summarize some of our notations. In \App{app:aux}, some auxiliary calculations are presented. Our paper also contains Supplementary Material \cite{foot:supp}. There, we overview the Weyl calculus on a curved configuration space, which is used in this work.

\section{General problem} 
\label{sec:gen}

Consider a wave propagating on an $n$-dimensional configuration space $M^n$ with coordinates $\vec{x} \equiv \{x^0, x^1, \ldots, x^{n-1}\}$ and some general metric tensor $g(\vec{x})$. For simplicity, assume that $M^n$ is diffeomorphic to $\mathbb{R}^n$, \ie the $n$-dimensional Euclidean space or pseudo-Euclidean space with the same metric signature as $M^n$. (For more information on why the diffeomorphism with $\mathbb{R}^n$ is needed, see \Sec{sec:xp}.) Suppose that the wave field $\field \equiv \field(\vec{x})$, which may have multiple components, is governed by a linear equation with no source terms,
\begin{gather}\label{eq:weq}
\oper{D} \field = 0,
\end{gather}
where $\oper{D}$ is a differential or, most generally, integral dispersion operator. (For vector waves, $\oper{D}$ is a matrix whose elements are operators; see \Sec{sec:vector}.) We shall assume that the GO parameter $\epsilon$ is small; namely,
\begin{gather}\label{eq:epsilon}
\epsilon \doteq \lambda/L \ll 1
\end{gather}
(the symbol $\doteq$ denotes definitions), where $\lambda$ is the characteristic wave period, or wavelength, and $L$ is the least characteristic scale among those of the wave envelope and of the medium, including the metric \cite{foot:GOappr}. Below, we propose a systematic reduction of \Eq{eq:weq} using the smallness of $\epsilon$ and eventually obtain a quasioptical model based on this equation. The idea of quasioptical modeling will be formalized later (Secs.~\ref{sec:qo} and \ref{sec:qoa}).

\new{As a side note, we emphasize that the theory to be developed does not describe wave transformations near cutoffs, where the GO parameter \eq{eq:epsilon} is not small. It is possible to waive this limitation, but formulating such generalized theory is left to future publications.}

\section{Envelope dispersion operator} 
\label{sec:env}

As the first step, let us introduce a unitary variable transformation
\begin{gather}\label{eq:u}
\field = \oper{\mc{U}}\envel, \quad \oper{\mc{U}} \doteq e^{i \theta (\vec{x})}.
\end{gather}
(Other $\oper{\mc{U}}$ may also be justified in some cases, \eg for dealing with caustics or quasiperiodic media \cite{my:qponder}, but we shall not consider this possibility in the present work.) The phase $\theta$, which we call the ``reference phase'', serves as a gauge potential. It is a real function such that
\begin{gather}\label{eq:kdef}
\vec{k}(\vec{x}) \doteq \nabla \theta(\vec{x})
\end{gather}
is the wave vector identical or close to that predicted by the GO approximation. This implies that the envelope $\envel$ and also $\vec{k}$ are slow functions [and that the wavelength entering \Eq{eq:epsilon} is $\lambda \approx 2\pi/k$]. Then, \Eq{eq:weq} becomes
\begin{gather}
\oper{\mc{D}} \envel = 0,\label{eq:Dtheta}
\end{gather}
where the ``envelope dispersion operator'' is $\oper{\mc{D}} \doteq \oper{\mc{U}}^\dag \oper{D} \oper{\mc{U}}$ (the dagger denotes the adjoint, as usual), or more explicitly,
\begin{gather}\label{eq:mcD0}
\oper{\mc{D}} = e^{-i \theta}\oper{D}\,e^{i \theta}.
\end{gather}

The reference phase $\theta$ is treated as a prescribed function. As will become clear later, knowing $\theta$ \textit{per~se} is not actually needed for our purposes; instead, it is $\vec{k}$ that matters. The latter can be calculated on some ``reference rays'' using the conventional ray equations \cite{book:tracy}
\begin{gather}\label{eq:rtp}
\frac{\dd x^\alpha}{\dd\tau} = \frac{\pd H}{\pd k_\alpha}, 
\quad
\frac{\dd k_\alpha}{\dd\tau} = -\frac{\pd H}{\pd x^\alpha}
\end{gather}
($\tau$ is any parameter along the ray), which also lead~to
\begin{gather}
H(\vec{x}, \vec{k}(\vec{x})) = \text{\rm const}.
\end{gather}
Note that, in general, $\theta$ and $\vec{k}$ are defined uniquely only to the leading order, so there exists some freedom in choosing reference rays and their Hamiltonian $H$. This means that more than one $\oper{\mc{D}}$ is possible. Still, envelope equations that (slightly) differ in the choice of $\vec{k}$ are equivalent in the sense that the \textit{total} field $\field$ that they describe is the same by construction. We shall discuss this in more detail in Secs.~\ref{sec:xgo} and \ref{sec:disc}.

Also note that our approach equally applies to stationary and nonstationary waves. In the case of a stationary wave, we assume that $\vec{x}$ is a coordinate in physical space (``spatial problem''), and the wave frequency $\omega$ serves as a constant parameter. In the case of a nonstationary wave, we assume that $\vec{x}$ is a coordinate in spacetime (``spacetime problem''), and then $\omega$ is a part of $\vec{k}$. In spacetime problems, we assume coordinates such that $x^0 = ct$, where $c$ is the speed of light, $t$ is time, and the metric signature is $(-,+,+,\ldots)$; then, $k_0 = - \omega/c$ (in case of the Minkowski metric, this implies that $k^0 = \omega/c$), so $\omega = - \pd_t \theta$, as usual. In other respects, spatial and spacetime problems are described on the same footing and will be distinguished only in \Sec{sec:EM}.

Having defined this terminology, we shall now discuss how $\oper{\mc{D}}$ can be expanded in $\epsilon$ asymptotically for any~$\oper{D}$.

\section{Scalar waves}
\label{sec:weylexp}

\new{Our asymptotic theory of mode-converting beams is based on the well-known phase-space methods in quantum and classical wave theory. (The key papers include, but are not limited to, \Refs{foot:bookweyl, ref:wigner32, ref:moyal49, ref:littlejohn86, ref:friedland87b, ref:mcdonald88, ref:littlejohn91}; for recent overviews, see \Refs{book:tracy, phd:ruiz17}.) In order to present our framework and notation in a self-contained manner, we restate some basics in \Sec{sec:notation} and Supplementary Material \cite{foot:supp}. Also note that our calculations in \Sec{sec:Dscalar} are similar in spirit to those in \Ref{ref:mcdonald88}, except we perform a higher-order expansion and assume a more general metric. We shall also comment throughout the paper on how our equations reproduce other known results in the corresponding limits.}

\subsection{Weyl calculus}
\label{sec:notation}

\subsubsection{Basic definitions}

Until \Sec{sec:vector}, we shall assume that $\field$ is a \textit{scalar} function. Any given operator $\oper{A}$ acting on it maps $\field$ to a new scalar function $\oper{A}\field$ that can be expressed as follows:
\begin{gather}\label{eq:A}
(\oper{A}\field)(\vec{x}) \doteq \int \dd^nx'\, \sqrt{\mdet{g}(\vec{x}')}\, \msf{A}(\vec{x}, \vec{x}') \field(\vec{x}').
\end{gather}
Here the integral is taken over $\mathbb{R}^n$ (and so are all integrals below, up to dimension), $\mdet{g} \doteq |\det g\, |$, and $\msf{A}$ is some kernel function that determines $\oper{A}$. Consider also a family of all unitary operators $\oper{A}_u$, which is a subset of all possible $\oper{A}$. For a given $\field$, all image functions $\oper{A}_u\field$ are mutually equivalent up to an isomorphism, so their family $\smash{\{\oper{A}_u\field\}}$ can be viewed as a single object, a ``state vector'' $\ket{\field}$, which belongs to a Hilbert space $\hilb{H}_1^n$ with inner product
\begin{gather} \label{eq:inner1}
\braket{\Psi | \Phi} \doteq \int \dd^nx \,\sqrt{\mdet{g}(\vec{x})}\, \Psi^*(\vec{x}) \Phi(\vec{x}). 
\end{gather}
Then, \Eq{eq:A} can be viewed as the ``$\vec{x}$ representation'' of $\oper{A}$, while the operator itself can be understood more generally as a transformation of $\ket{\field}$, \ie of the whole family $\smash{\{\oper{A}_u\field\}}$. Using this invariant notation, one can formulate a machinery, called the Weyl calculus \cite{foot:wc}, that allows efficient asymptotic approximation of operators using the smallness of $\epsilon$. Below, we overview the key theorems of the Weyl calculus that are used in our paper. Readers who are interested in details and proofs of these theorems can find them in the Supplementary Material~\cite{foot:supp}.

\subsubsection{Coordinate and momentum operators}
\label{sec:xp}

We start by defining the coordinate and momentum (wave-vector) operators
\begin{gather}
\boper{x} = \{\oper{x}^0, \oper{x}^1, \ldots, \oper{x}^{n-1}\}, 
\\
\boper{p} = \{\oper{p}_0, \oper{p}_1, \ldots, \oper{p}_{n-1}\}
\end{gather}
such that the $\vec{x}$~representations of  $\oper{x}^\mu$ and $\oper{p}_\mu$ be as follows:
\begin{gather}
\oper{x}^\mu\field = x^\mu\field, \label{eq:sxp1}\\
\oper{p}_\mu \field = -i \mdet{g}^{-1/4} \pd_\mu (\mdet{g}^{1/4} \field).\label{eq:sxp2}
\end{gather}
Here $\pd_\mu \doteq \pd/\pd x^\mu,$ $x^\mu$ and $p_\mu$ are the corresponding eigenvalues, and the factors $\smash{\mdet{g}^{\pm 1/4}}$ are introduced to keep $\boper{p}$ self-adjoint under the inner product \eq{eq:inner1} \cite{ref:dewitt52}. (This would not be the case if $M^n$ were not diffeomorphic to $\mathbb{R}^n$ \cite{ref:domingos84}.) Since
\begin{gather}\label{eq:qdef}
\oper{p}_\nu = -i\pd_\nu - iq_\nu(\vec{x}), \quad q_\nu \doteq \frac{1}{4}\,\pd_\nu (\ln \mdet{g}),
\end{gather}
and $q_\nu$ commutes with $\oper{x}^\mu$, one arrives at the usual commutation relation $[\oper{x}^\mu, \oper{p}_\nu] = i\delta^\mu_\nu$. 

Let us consider the eigenvectors $\ket{\vec{x}}$ and $\ket{\vec{p}}$ of the coordinate and momentum operators, which are defined~as
\begin{gather}
\boper{x}\ket{\vec{x}} = \vec{x}\ket{\vec{x}},
\quad
\boper{p}\ket{\vec{p}} = \vec{p}\ket{\vec{p}}.
\end{gather}
Since the operators are self-adjoint, these eigenvectors can be chosen as mutually orthogonal, and we shall assume the following normalization:
\begin{gather}
\braket{\vec{x}_1 | \vec{x}_2} = \mcc{G}(\vec{x}_1, \vec{x}_2)\,\delta(\vec{x}_1 - \vec{x}_2), \label{eq:xnorm}\\
\braket{\vec{p}_1 | \vec{p}_2} = \bar{\mcc{G}}(\vec{p}_1, \vec{p}_2)\,\delta(\vec{p}_1 - \vec{p}_2). \label{eq:pnorm}
\end{gather}
Here we introduced
\begin{gather}
\mcc{G}(\vec{x}_1, \vec{x}_2) \doteq [\mdet{g}(\vec{x}_1)\mdet{g}(\vec{x}_2)]^{-1/4},\\
\bar{\mcc{G}}(\vec{p}_1, \vec{p}_2) \doteq [\mdet{\bar{g}}(\vec{p}_1)\mdet{\bar{g}}(\vec{p}_2)]^{-1/4}.
\end{gather}
The function $\mdet{\bar{g}}$ can be chosen arbitrarily as long as it is kept positive. It plays a role similar to that of $\mdet{g}$ in the Weyl calculus, but note that this is just a normalization factor, and we introduced it only to maintain the symmetry between $\boper{x}$ and $\boper{p}$. One can show then \cite{foot:supp} that our original function $\field$ and the envelope $\envel$ can be expressed through the corresponding state vectors~as
\begin{gather}
 \field(\vec{x}) = \braket{\vec{x} | \field}, 
 \quad
 \envel(\vec{x}) = \braket{\vec{x} | \envel}.
\end{gather}
The $\vec{p}$ representations of these state vectors are introduced similarly as $\braket{\vec{p} | \field}$ and $\braket{\vec{p} | \envel}$. One can also show \cite{foot:supp} that $\braket{\vec{x}_1 | \oper{A} | \vec{x}_2} = \msf{A}(\vec{x}_1, \vec{x}_2)$, and
\begin{gather}\label{eq:cdot}
\braket{\vec{x} | \vec{p}} = \braket{\vec{p} | \vec{x}}^* = 
\frac{\exp(i\vec{p}\cdot\vec{x}),}{(2\pi)^{n/2} [\mdet{g}(\vec{x}) \mdet{\bar{g}}(\vec{p})]^{1/4}}.
\end{gather}
Here, $\vec{p}\cdot\vec{x} \doteq p_\mu x^\mu$, and summation over repeated indices is assumed here and further.

\subsubsection{Wigner--Weyl transform}
\label{sec:wwt}

The set of all eigenvalues of the coordinate and momentum operators form a $2n$-dimensional ``phase space'' $z \equiv (\vec{x}, \vec{p})$. For every given $z$, we introduce the so-called \textit{Wigner operator} $\oper{\Delta}_z$, which is self-adjoint and defined as
\begin{gather}
\oper{\Delta}_z \doteq \int \dd^ns\, G(\vec{x}, \vec{s})\,
\ket{\vec{x}-\vec{s}/2}
\bra{\vec{x}+\vec{s}/2} e^{-i \vec{p}\cdot \vec{s}},\\
G(\vec{x}, \vec{s}) \doteq [\mdet{g}(\vec{x} - \vec{s}/2) \mdet{g}(\vec{x} + \vec{s}/2)]^{1/4}.\label{eq:G}
\end{gather}
(Note that $G = 1$ if the $\vec{x}$ space is Euclidean or pseudo-Euclidean.) Using $\oper{\Delta}_z$, we define the \textit{Wigner--Weyl transform} $\weyl_z: \oper{A} \mapsto A$, which maps a given operator $\oper{A}$ on $\hilb{H}_1^n$ to a function $A$ (``Weyl symbol'') on the $z$ space. Specifically, the Weyl symbol of a given operator $\oper{A}$ is $A(\vec{x},\vec{p})\doteq \text{tr}(\oper{\Delta}_z\oper{A})$ (``tr'' stands for trace); \ie
\begin{gather}\label{eq:Aw}
A(\vec{x},\vec{p}) \doteq \int \dd^ns\,G(\vec{x}, \vec{s})\,
\braket{\vec{x}+\vec{s}/2 | \oper{A} | \vec{x}-\vec{s}/2}e^{-i \vec{p}\cdot \vec{s}}.
\end{gather}

As can be seen easily from this definition, if a given operator is self-adjoint, then its Weyl symbol is real. This also leads to the following corollary. Consider splitting a given $\oper{A}$ as $\oper{A} = \oper{A}_H + i\oper{A}_A$, where the subscripts denote the Hermitian and anti-Hermitian parts,
\begin{gather}\label{eq:aux3003}
\oper{A}_H \doteq \frac{1}{2}\,(\oper{A} + \oper{A}^\dag),
\quad
\oper{A}_A \doteq \frac{1}{2i}\,(\oper{A} - \oper{A}^\dag).
\end{gather}
Both $\oper{A}_H$ and $\oper{A}_A$ (not to be confused with $i\oper{A}_A$) are self-adjoint by definition. Thus, the corresponding Weyl images $A_H$ and $A_A$ are real. 

We also define the \textit{inverse Wigner--Weyl transform} $\weyl^{-1}: A \mapsto \oper{A}$, which maps a given function $A$ to the corresponding operator $\oper{A}$ via
\begin{gather}\label{eq:invweyl}
\oper{A} 
= \frac{1}{(2\pi)^n}\int \dd^nx\,\dd^np\, A(\vec{x},\vec{p}) \oper{\Delta}_z.
\end{gather}
The direct and inverse transforms set the ``Weyl correspondence'' between operators and functions on the $(\vec{x},\vec{p})$ space, $\oper{A} \Leftrightarrow  A(\vec{x},\vec{p})$. As can be checked by a direct calculation, for any function $f$, one has
\begin{gather}\label{eq:wcor}
f(\boper{x}) \Leftrightarrow f(\vec{x}), \quad f(\boper{p}) \Leftrightarrow f(\vec{p}).
\end{gather}
However, the Weyl symbols of operators that cannot be represented as $f_1(\boper{x}) + f_2(\boper{p})$ are generally more complicated. In particular, one can show that \cite{foot:supp}
\begin{gather}
f(\boper{x})\oper{p}_\alpha
\kern 5pt \Leftrightarrow \kern 5pt
p_\alpha f(\vec{x}) + \frac{i}{2}\,\pd_\alpha f(\vec{x}),\\
\oper{p}_\alpha f(\boper{x}) 
\kern 5pt \Leftrightarrow \kern 5pt
p_\alpha f(\vec{x}) - \frac{i}{2}\,\pd_\alpha f(\vec{x}),
\end{gather}
and also \cite{foot:supp}
\begin{align}\label{eq:TTT}
\oper{p}_{\alpha}f(\oper{\vec{x}}) \oper{p}_{\beta} 
\quad \Leftrightarrow \quad & p_{\alpha} p_{\beta}f(\vec{x})
+\frac{1}{4}\,\pd_{\alpha \beta}^2f(\vec{x})  \notag\\
& +\frac{i}{2}\,p_{\alpha}\pd_{\beta}f(\vec{x})
-\frac{i}{2}\, p_{\beta}\pd_{\alpha}f(\vec{x}).
\end{align}
Overall, the Weyl symbol of an operator that is \textit{any} given combination $f(\boper{x}, \boper{p})$ of $\boper{x}$ and $\boper{p}$ approaches $f(\vec{x}, \vec{p})$ in the GO limit, when $[\oper{x}^\mu, \oper{p}_\nu]$ is negligible. However, in general, $f(\boper{x}, \boper{p})$ does not map simply to $f(\vec{x}, \vec{p})$.

\subsection{Approximate $\boldsymbol{\oper{\mc{D}}}$}
\label{sec:Dscalar}

Using the notation introduced in \Sec{sec:notation}, one can express \Eq{eq:weq} in the following invariant form:
\begin{gather}\label{eq:weq1}
\oper{D} \ket{\field} = 0.
\end{gather}
Accordingly, \Eq{eq:Dtheta} becomes
\begin{gather}\label{eq:Dpsienvel}
\oper{\mc{D}} \ket{\envel} = 0,
\end{gather}
where the invariant form of the envelope dispersion operator [\Eq{eq:mcD0}] is as follows:
\begin{gather}\label{eq:mcD}
\oper{\mc{D}} = e^{-i \theta (\boper{x})}\oper{D}\,e^{i \theta (\boper{x})}.
\end{gather}

We shall now approximate $\oper{\mc{D}}$ in three steps: (i) we map the right-hand side of \Eq{eq:mcD} onto a function, or symbol, using the Wigner--Weyl transform; (ii) we approximate this symbol using the smallness of $\epsilon$ [\Eq{eq:epsilon}]; and (iii) we produce an operator out of the approximated symbol using the inverse Wigner--Weyl transform.

\new{The first two steps can be done by expressing the Weyl symbol of $\oper{\mc{D}}$ as $\mc{D}(\vec{x}, \vec{p}) = e^{-i \theta (\vec{x})} \star D(\vec{x}, \vec{p}) \star e^{i \theta (\vec{x})}$, where $D$ is the Weyl symbol of $\oper{D}$ and $\star$ is the Moyal product \cite{foot:supp}. By expanding $\star$ in the GO parameter, one can then obtain an approximate symbol of $\oper{\mc{D}}$ formally. But here we adopt a different approach, which is more transparent. We start by expressing $\oper{D}$ through $D$ explicitly,}
\begin{widetext}
\begin{gather}
\oper{D}=\frac{1}{(2\pi)^n}\int \dd^nx\,\dd^np\,\dd^ns\, G(\vec{x}, \vec{s})\,
\ket{\vec{x}-\vec{s}/2}e^{-i \vec{p}\cdot \vec{s}}D(\vec{x}, \vec{p})
\bra{\vec{x}+\vec{s}/2},
\end{gather}
where $G$ is a metric factor given by \Eq{eq:G}. Using the fact that $\theta (\boper{x})\ket{\vec{x} \pm \vec{s}/2} = \theta(\vec{x} \pm \vec{s}/2)\ket{\vec{x} \pm \vec{s}/2}$, one can rewrite \Eq{eq:mcD} as follows:
\begin{gather}
\oper{\mc{D}}
=\frac{1}{(2\pi)^n}\int \dd^nx\,\dd^np\,\dd^ns\,G(\vec{x}, \vec{s})\,
e^{i [\theta (\vec{x}+\vec{s}/2)-\theta (\vec{x}-\vec{s}/2)-\vec{p}\cdot \vec{s}]}
\ket{\vec{x}-\vec{s}/2}
D(\vec{x},\vec{p})\bra{\vec{x}+\vec{s}/2}.
\end{gather}
Consider a formal Taylor expansion of the reference phase $\theta$ in $\vec{s}$:
\begin{gather}
\theta(\vec{x} \pm \vec{s}/2)
= \theta(\vec{x}) 
\pm \frac{s^{\alpha}}{2}\frac{\pd \theta (\vec{x})}{\pd x^{\alpha}}
+ \frac{1}{2}\frac{s^{\alpha}s^{\beta}}{4}\frac{\pd^2\theta (\vec{x})}{\pd x^{\alpha}\pd x^{\beta}}
\pm \frac{1}{6}\frac{s^{\alpha}s^{\beta}s^{\gamma}}{8}\frac{\pd^3\theta (\vec{x})}{\pd x^{\alpha}\pd x^{\beta}\pd x^{\gamma}}
+\ldots
\end{gather}
This gives
\begin{gather}
\theta(\vec{x}+\vec{s}/2) - \theta(\vec{x}-\vec{s}/2) = 
s^{\alpha}\frac{\pd \theta (\vec{x})}{\pd x^{\alpha}}
+\frac{s^{\alpha}s^{\beta}s^{\gamma}}{24}\frac{\pd^3\theta(\vec{x})}{\pd x^{\alpha}\pd x^{\beta}\pd x^{\gamma}}
+\ldots
=\vec{k}(\vec{x})\cdot \vec{s}+\frac{s^{\alpha}s^{\beta}s^{\gamma}}{24}\frac{\pd^2k_\gamma(\vec{x})}{\pd x^{\alpha}\pd x^{\beta}}+\ldots,
\end{gather}
so
\begin{align}
e^{i [\theta (\vec{x}+\vec{s}/2)-\theta (\vec{x}-\vec{s}/2)-\vec{p}\cdot \vec{s}]}
& =\left[1+\frac{i}{24}s^{\alpha}s^{\beta}s^{\gamma}\frac{\pd^2k_\gamma(\vec{x})}{\pd x^{\alpha}\pd x^{\beta}}+\ldots\right] 
e^{i[\vec{k}(\vec{x})-\vec{p}]\cdot \vec{s}} \notag\\
& =\left[1+\frac{1}{24}\frac{\pd^2k_\gamma(\vec{x})}{\pd x^{\alpha}\pd x^{\beta}}
\frac{\pd^3}{\pd p_{\alpha}\pd p_{\beta}\pd p_{\gamma}}+\ldots\right]e^{i [\vec{k}(\vec{x})-\vec{p}]\cdot \vec{s}}.
\label{eq:aux1}
\end{align}
This leads to the following expression for the envelope dispersion operator:
\begin{align}
\oper{\mc{D}}
& =\frac{1}{(2\pi)^n}\int \dd^nx\,\dd^np\,\dd^ns\,G(\vec{x}, \vec{s})\,
\left\{\left[1+\frac{1}{24}\frac{\pd^2k_\gamma(\vec{x})}{\pd x^{\alpha}\pd x^{\beta}}
\frac{\pd^3}{\pd p_{\alpha}\pd p_{\beta}\pd p_{\gamma}}+\ldots\right]
e^{i [\vec{k}(\vec{x})-\vec{p}]\cdot \vec{s}}\right\}
\ket{\vec{x}-\vec{s}/2}D(\vec{x},\vec{p})\bra{\vec{x}+\vec{s}/2}\notag\\
& =\frac{1}{(2\pi)^n}\int \dd^nx\,\dd^np\,\dd^ns\,G(\vec{x}, \vec{s})\,e^{i [\vec{k}(\vec{x})-\vec{p}]\cdot \vec{s}}
\ket{\vec{x}-\vec{s}/2}
\left[D(\vec{x},\vec{p})
-\frac{1}{24}\frac{\pd^2k_\gamma(\vec{x})}{\pd x^{\alpha}\pd x^{\beta}}
\frac{\pd^3D(\vec{x},\vec{p})}{\pd p_{\alpha}\pd p_{\beta}\pd p_{\gamma}}
+\ldots\right]
\bra{\vec{x}+\vec{s}/2}\notag\\
& = \frac{1}{(2\pi)^n}\int \dd^nx\,\dd^np\,\dd^ns\,G(\vec{x}, \vec{s})\,
e^{-i \vec{p}\cdot \vec{s}}
\ket{\vec{x}-\vec{s}/2} D' \boldparen{\vec{x},\vec{k}(\vec{x})+\vec{p}} \bra{\vec{x}+\vec{s}/2},
\end{align}
where we integrated by parts and introduced an ``effective dispersion function''
\begin{gather}\label{eq:Deff}
D'(\vec{x},\vec{p}) \doteq D(\vec{x},\vec{p})
-\frac{1}{24}\frac{\pd^2k_\gamma(\vec{x})}{\pd x^{\alpha}\pd x^{\beta}}
\frac{\pd^3D(\vec{x},\vec{p})}{\pd p_{\alpha}\pd p_{\beta}\pd p_{\gamma}}
+\ldots
\end{gather}
\new{The first term on the right-hand side of \Eq{eq:Deff} is $O(1)$, because at $\epsilon \to 0$, it becomes the GO dispersion function, which is assumed order-one \cite{foot:orderone}. (At nonzero $\epsilon$, the Weyl symbol $D$ generally does \textit{not} coincide with the dispersion function that governs the wave in a homogeneous medium; see \Sec{sec:wwt}.) The second term can be estimated by assuming $\pd/\pd x \sim 1/L$ and $\pd/\pd p \sim 1/k$, since we are interested in $D'(\vec{x}, \vec{p})$ at $\vec{p}$ close to $\vec{k}$. Then,
\begin{gather}
\frac{\pd^2k_\gamma}{\pd x^{\alpha}\pd x^{\beta}}
\frac{\pd^3D}{\pd p_{\alpha}\pd p_{\beta}\pd p_{\gamma}}
\sim \frac{k}{L^2}\,\frac{D}{k^3} 
\sim \epsilon^2 D,
\end{gather}
meaning that the second term in \Eq{eq:Deff} is roughly $\epsilon^2$ times smaller than the first one. This fact will be used below.}

The inverse Wigner--Weyl transform will turn the coordinate $\vec{p}$ into the operator $\boper{p}$. The latter will act on the wave envelope [\Eq{eq:Dpsienvel}], which is considered slow in the coordinate representation, so $\boper{p}\ket{\envel} = O(\epsilon)$. In this sense, $p = O(\epsilon)$, so by Taylor-expanding $D'$ in $\vec{p}$, we are effectively expanding $\oper{\mc{D}}$ in $\epsilon$. It is sufficient for our purposes to adopt the second-order expansion,
\begin{gather}
\oper{\mc{D}} \approx \frac{1}{(2\pi)^n}\int \dd^nx\,\dd^np\,\dd^ns\,G(\vec{x}, \vec{s})\,
e^{-i \vec{p}\cdot \vec{s}}
\ket{\vec{x}-\vec{s}/2}
\Big[
D'(\vec{x}) + p_{\mu}\mc{V}^{\mu}(\vec{x})
+ \frac{1}{2}\,p_{\mu}p_{\nu}\Theta^{\mu\nu}(\vec{x})\Big]
\bra{\vec{x}+\vec{s}/2},
\end{gather}
\end{widetext}
where we introduced $D'(\vec{x}) \doteq D'\boldparen{\vec{x},\vec{k}(\vec{x})}$ and
\begin{gather}
\mc{V}^{\mu}(\vec{x}) \doteq \frac{\pd D' \boldparen{\vec{x},\vec{k}(\vec{x})}}{\pd k_{\mu}},
\label{eq:aux45010}\\
\Theta^{\mu \nu}(\vec{x}) \doteq \frac{\pd^2 D' \boldparen{\vec{x},\vec{k}(\vec{x})}}{\pd k_{\mu}\pd k_{\nu}}.
\label{eq:aux45020}
\end{gather}

By properties of the inverse Wigner--Weyl transform $\weyl^{-1}$ (\Sec{sec:wwt}), $\oper{\mc{D}}$ can be written as follows:
\begin{multline}\label{eq:aux1001}
\oper{\mc{D}} \approx  \weyl^{-1}[D'(\vec{x})]
+\weyl^{-1}\left[p_{\mu}\mc{V}^\mu(\vec{x})\right]
\\
+\frac{1}{2}\,\weyl^{-1}\left[p_{\mu}p_{\nu}\Theta^{\mu \nu}(\vec{x})\right].
\end{multline}
The inverse Wigner--Weyl transforms here can be calculated using \Eqs{eq:wcor}-\eq{eq:TTT} and $\Theta^{\mu\nu} = \Theta^{\nu\mu}$. This leads~to
\begin{multline}\label{eq:mcdscal}
\oper{\mc{D}} \approx D'(\boper{x})
+ \frac{1}{2}\left[\oper{p}_\mu \mc{V}^\mu(\boper{x})
+\mc{V}^\mu(\boper{x}) \oper{p}_\mu\right]
\\+ \frac{1}{2}\,\oper{p}_{\mu}\Theta^{\mu \nu}(\boper{x})\oper{p}_{\nu}
-\frac{1}{8}\,(\pd_{\mu\nu}^2\Theta^{\mu \nu})(\boper{x}).
\end{multline}

\new{One can further simplify \Eq{eq:mcdscal} as follows. Suppose that the medium, including the metric, is characterized by some parameters $\msf{P}$. Within the accuracy of the theory developed in this paper, we shall neglect terms involving  $\pd^2\msf{P}$ and $(\pd\msf{P})(\pd\psi)$ while generally retaining terms such as $\pd^2\envel$ (as to be explained \Sec{sec:qoe}), where $\pd$ denotes a generic spatial derivative. Hence, we can ignore the difference between $D$ and $D'$, and we can also omit $\smash{\pd_{\mu\nu}^2\Theta^{\mu \nu}}$ in \Eq{eq:mcdscal}, so $\oper{\mc{D}}$ can be expressed as follows:
\begin{gather}\label{eq:mcdscal2}
\oper{\mc{D}} \approx D(\boper{x})
+ \frac{1}{2}\left[\oper{p}_\mu \mc{V}^\mu(\boper{x})
+\mc{V}^\mu(\boper{x}) \oper{p}_\mu\right]
+ \frac{1}{2}\,\oper{p}_{\mu}\Theta^{\mu \nu}(\boper{x})\oper{p}_{\nu}.
\end{gather}
Also, $\mc{V}^\mu$ and $\Theta^{\mu \nu}$ can be calculated through the zeroth-$\epsilon$ limit of $D$, denoted $D_0$, which is just the dispersion function of the homogeneous medium:
\begin{gather}\notag
\mc{V}^{\mu}(\vec{x}) \approx 
\frac{\pd D_0 \boldparen{\vec{x},\vec{k}(\vec{x})}}{\pd k_{\mu}}
,
\quad
\Theta^{\mu \nu}(\vec{x}) 
\approx \frac{\pd^2 D_0 \boldparen{\vec{x},\vec{k}(\vec{x})}}{\pd k_{\mu}\pd k_{\nu}}.
\end{gather}
Importantly, $D_0(\vec{x},\vec{k})$ can depend only on $\vec{k}$ and on local parameters of the medium (but not on their derivatives). As a gradient of a scalar [\Eq{eq:kdef}], $\vec{k}$ is a true covector; hence, $D_0(\vec{x},\vec{k})$ is a true scalar. This makes $\mc{V}^{\mu}$ a vector and $\Theta^{\mu\nu}$ a tensor within the assumed accuracy.}

\new{As a side remark, note that vectors like $\mc{V}^\mu$, which are denoted with upper indices, belong to the space tangent to $M^n$. (Such vectors must not be confused with multi-component fields denoted with Latin indices in \Sec{sec:vector}.) Similarly, covectors like $k_\mu$, which are denoted with lower indices, belong to the space cotangent to $M^n$. For any given $X$, one does not have to distinguish $X^\mu$ and $X_\mu$ only if the metric is Euclidean. In the case of a general metric $g_{\mu\nu}$, one has
\begin{gather}
X_\mu = g_{\mu\nu} X^\nu, \quad X^\mu = g^{\mu\nu} X_\nu.
\end{gather}
where the matrix $g^{\mu\nu}$ is the inverse of $g_{\mu\nu}$. Similar rules apply to manipulating tensors, as usual.}

\subsection{Approximate envelope equation}
\label{sec:aenv}

Let us derive the $\vec{x}$~representation of \Eq{eq:mcdscal2}, which we need to obtain an explicit form of \Eq{eq:Dtheta}. From the $\vec{x}$~representation of $\boper{x}$ and $\boper{p}$ given by \Eqs{eq:sxp1} and \eq{eq:sxp2}, terms like $\boper{p} f(\boper{x})$ (for any $f$) must be interpreted as
\begin{gather}
\left[\oper{p}_\mu f(\boper{x})\right]\envel 
=-i\mdet{g}^{-1/4} \pd_\mu [\mdet{g}^{1/4} f(\vec{x}) \envel],
\end{gather}
where $g$ and $\envel$ are also functions of $\vec{x}$. Then, 
\new{
\begin{multline}\label{eq:aux4503} 
\oper{\mc{D}} \envel \approx D \envel
- i \mc{V}^\mu \pd_\mu \envel - \frac{i}{2}\,\mc{V}^\mu{}_{;\mu} \envel
\\- \frac{1}{2\mdet{g}^{1/4}}\,\pd_\mu[\Theta^{\mu \nu} \pd_\nu (\mdet{g}^{1/4}\envel)],
\end{multline}
where all terms are functions of $\vec{x}$, and $_{;\mu}$ denotes the covariant derivative with respect to~$x^\mu$, so $X^\mu{}_{;\mu}$ (for any given vector $X^\mu$) is the divergence; namely,
\begin{gather}\label{eq:div}
X^\mu{}_{;\mu} \doteq \frac{1}{\sqrt{\mdet{g}}}\,\frac{\pd}{\pd x^\mu}\,(\sqrt{\mdet{g}} X^\mu).
\end{gather}
}

\new{Except for $D \envel$, all terms in \Eq{eq:aux4503} are covariant (\ie invariant with respect to coordinate transformations) within the accuracy of our theory. Hence, the symbol $D$ is also covariant within the accuracy of our theory [this includes not just the zeroth-order dispersion function $D_0$ but also the leading-order correction $D - D_0  = O(\pd\msf{P})$ that may appear in inhomogeneous medium], even though this may not be obvious directly from the definition of the Weyl transform (\Sec{sec:notation}). To emphasize this fact, and also to ensure consistency of notation with the vector-wave theory to be discussed in the later sections, we denote this covariant function as~$\Deff$, and we also adopt the notation
\begin{gather}
\Deffr \doteq \text{Re}\,\Deff,
\quad
\Deffi \doteq \text{Im}\,\Deff.
\end{gather}
Then, the approximate form of the envelope equation \eq{eq:Dtheta} can be summarized as follows:
\begin{gather}\label{eq:DHDA}
(\oper{\mc{D}}_H + i\oper{\mc{D}}_A)\psi = 0.
\end{gather}
Here, $\oper{\mc{D}}_H$ and $\oper{\mc{D}}_A$ are, respectively, the Hermitian and anti-Hermitian parts of the approximated $\oper{\mc{D}}$, namely,
\begin{multline}\label{eq:dprime}
\oper{\mc{D}}_H \envel = \Deffr\envel
- i V^\mu \,\envel_{,\mu} - \frac{i}{2}\,V^\mu{}_{;\mu} \envel
\\- \frac{1}{2\mdet{g}^{1/4}}\,[\Deffr^{|\mu\nu} (\mdet{g}^{1/4}\envel)_{,\nu}]_{,\mu},
\end{multline}
\begin{multline}\label{eq:ddprime}
\oper{\mc{D}}_A\envel = \Deffi \envel
-i\mc{V}_A^\mu \,\envel_{,\mu} - \frac{i}{2}\,\mc{V}_A^\mu{}_{;\mu} \envel
\\- \frac{1}{2\mdet{g}^{1/4}}\,[\Deffi^{|\mu\nu} (\mdet{g}^{1/4}\envel)_{,\nu}]_{,\mu}
\end{multline}
[from now on, we do not emphasize the approximate nature of \Eq{eq:aux4503}, so $\approx$ will be replaced with~$=$], and
\begin{gather}\label{eq:VP}
V^{\mu} \doteq \mc{V}_H^{\mu} = \Deffr^{|\mu},
\end{gather}
where we assume a standard notation}
\begin{gather}
f_{,\mu} \doteq \frac{\pd f}{\pd x^\mu},
\quad
f_{,\mu\nu} \doteq \frac{\pd^2 f}{\pd x^\mu \pd x^\nu}.
\end{gather}
Note that $_{,\mu}$ is the ``full'' derivative in the sense that, for any given $f\boldparen{\vec{x}, \vec{k}(\vec{x})}$, it applies both to the first and the second argument of $f$; namely, the chain rule leads to
\begin{gather}\label{eq:comma}
f_{,\mu}\boldparen{\vec{x}, \vec{k}(\vec{x})} = f_{|\mu} + f^{|\nu} k_{\nu|\mu}.
\end{gather}
The true partial derivatives are introduced as follows:
\begin{gather}
f_{|\mu} \doteq \frac{\pd f(\vec{x}, \vec{k})}{\pd x^\mu},
\quad
f_{|\mu\nu} \doteq \frac{\pd^2 f(\vec{x}, \vec{k})}{\pd x^\mu\pd x^\nu},\\
f^{|\mu} \doteq \frac{\pd f(\vec{x}, \vec{k})}{\pd k_\mu},
\quad
f^{|\mu\nu} \doteq \frac{\pd^2 f(\vec{x}, \vec{k})}{\pd k_\mu\pd k_\nu}.
\end{gather}
The derivatives $_{,\mu}$ and $_{|\mu}$ are equivalent for functions that depend only on $\vec{x}$. In particular, ${k_\nu \equiv \pd_\nu \theta(\vec{x})}$ satisfies
\begin{gather}
k_{\nu|\mu} = k_{\nu,\mu} = k_{\mu,\nu} 
= \theta_{,\mu\nu} = \theta_{,\nu\mu}.
\end{gather}

\subsection{Leading-order approximation:\\geometrical optics}
\label{sec:first}

For the envelope approximation to hold, \ie for $\envel(\vec{x})$ to remain a slow function, $\vec{k}(\vec{x})$ must be chosen such that $\Deff\boldparen{\vec{x}, \vec{k}(\vec{x})} \envel \lesssim O(\epsilon)$. [The other terms in \Eq{eq:Dtheta} are automatically small on the account of \Eq{eq:epsilon}.] Let us adopt the usual GO ordering \cite{foot:dissip}
\begin{gather}\label{eq:ordering} 
 \Deffr = O(1), \quad \Deffi = O(\epsilon),
\end{gather}
or more rigorously, $\Deffi \lesssim O(\epsilon)$; \ie $\Deffi$ much smaller than $O(\epsilon)$ is also allowed \cite{foot:herm}. Then, one can define $\vec{k}(\vec{x})$ such that 
\begin{gather}\label{eq:drOm}
\Deffr\boldparen{\vec{x}, \vec{k}(\vec{x})} = 0.
\end{gather}
This can be achieved by calculating $\vec{k}$ as discussed in \Sec{sec:env} with the ray Hamiltonian ${H = \Deffr}$ and the initial condition [at any chosen $\smash{\vec{x}^{(0)}}$ and $\smash{\vec{k}^{(0)}}$] such that
\begin{gather}
\Deffr\boldparen{\vec{x}^{(0)}, \vec{k}^{(0)}} = 0.
\end{gather} 

To the first order in $\epsilon$, \Eq{eq:Dtheta} is
\begin{gather}\label{eq:gopsi}
V^\mu \,\envel_{,\mu} + \frac{1}{2}\,V^\mu{}_{;\mu} \envel = \Deffi \envel,
\end{gather}
and as a corollary,
\begin{gather}\label{eq:actgo}
\mc{J}^\mu{}_{;\mu} = -2\Deffi |\envel|^2, \quad \mc{J}^\mu \doteq -V^\mu |\envel|^2.
\end{gather}
At zero $\Deffi$, the dynamics is conservative, and \Eq{eq:actgo} reflects conservation of the wave action, or quanta \cite{my:amc}. Accordingly, $\Deffi$ determines the dissipation rate, $\mc{J}^\mu$ can be identified (at least, up to a constant factor) as the action flux density, and $V^\mu$ is proportional to the group velocity (in space or in spacetime, depending on the problem). Equations \eq{eq:drOm}-\eq{eq:actgo} coincide with the equations of the traditional GO theory \cite{book:tracy, my:amc}. Below, we extend them by retaining the second-order terms neglected in~\Eq{eq:gopsi}.

\subsection{Quasioptical model} 
\label{sec:qo}

\subsubsection{Ray-based coordinates}
\label{sec:metric}

Let us start by introducing ray-based coordinates as follows \cite{foot:mymetric}. Suppose multiple rays launched with different $\vec{x}^{(0)}$ within the beam. Their trajectories can be found using the ray equations as discussed in \Sec{sec:first}; this determines the path $\zeta$ along each ray as a function of the initial coordinate. We treat $\zeta$ as the longitudinal coordinate along the wave beam, and each of its isosurfaces is considered as the transverse space $M^{n-1}_\perp(\zeta)$. The coordinates on this space, $\vec{\varrho} \equiv \{\varrho^1, \varrho^2, \ldots, \varrho^{n-1}\}$, can be introduced arbitrarily (yet they will be specified below). Then, we define $n-1$ independent vector fields $\vec{e}_\sigma$ via
\begin{gather}
\vec{e}_\sigma \doteq \pd \vec{x}/\pd \varrho^\sigma
\end{gather}
and another vector $\vec{e}_0$ such that it is linearly-independent from all $\vec{e}_\sigma$. Hence, any $\dd\vec{x}$ can be decomposed as follows:
\begin{gather}\label{eq:dx}
\dd\vec{x} = \vec{e}_0 \dd\zeta + \vec{e}_\sigma \dd\varrho^\sigma,
\end{gather}
and summation over repeated indices $\sigma$ (and $\sigmap$) is henceforth assumed from 1 to $n-1$. 

Let us also construct the dual basis $\{\vec{e}^\mu\}$. (We treat vectors and covectors on the same footing; namely, $\vec{X} \cdot \vec{Y} = X_\mu Y^\mu = X^\mu Y_\mu$, and $\vec{X} = \vec{Y}$ equally means $X^\mu = Y^\mu$ and $X_\mu = Y_\mu$.) By definition,
\begin{gather}\label{eq:ort}
\vec{e}^\mu \cdot \vec{e}_\nu = \delta^\mu_\nu,
\end{gather}
so we adopt $\vec{e}^0 \doteq \del \zeta$. Then, the general form of $\vec{e}_0$ is
\begin{gather}
\vec{e}_0 = \pm \alpha^2\vec{e}^0 + \vec{\beta},
\end{gather}
where the first sign is determined by the metric signature, $\alpha \doteq |\vec{e}^0 \cdot \vec{e}^0|^{-1/2}$, $\vec{\beta} \doteq \vec{e}_\sigma \beta^\sigma$, and $\beta^\sigma$ are arbitrary coefficients. This leads to the following metric representation in the coordinates $\{\zeta, \vec{\varrho}\}$:
\begin{gather}
g = \left(
\begin{array}{cc}
\pm \alpha^2 +  \vec{\beta}^\intercal \boldsymbol{h} \vec{\beta} & (\boldsymbol{h} \vec{\beta})^\intercal\\[5pt]
\boldsymbol{h} \vec{\beta} & \boldsymbol{h}
\end{array}
\right),
\end{gather}
Here, $\boldsymbol{h}$ is the matrix with elements $h_{\sigma\sigmap} \doteq \vec{e}_\sigma \cdot \vec{e}_{\sigmap}$, and $^\intercal$ denotes transposition. By using a known theorem for the determinant of a block matrix \cite{foot:det}, one obtains
\begin{gather}
\mdet{g} = \alpha^2\mdet{h}, \quad \mdet{h} \doteq |\det \boldsymbol{h}\,|.
\end{gather}
Let us also introduce the transverse projection of $\dd\vec{\varrho}$,
\begin{gather}
\dd\vec{\varrho}_\perp \doteq (\mathbb{1} \mp \alpha^2 \vec{e}^0\vec{e}^0) \dd\vec{\varrho},
\quad
\dd\varrho_\perp^\sigma = \dd\varrho^\sigma + \beta^\sigma \dd\zeta,
\end{gather}
where $\mathbb{1}$ is a unit matrix and $\vec{e}^0\vec{e}^0$ is a dyad formed out of $\vec{e}^0$. Then, one obtains from \Eq{eq:dx}~that
\begin{gather}
\dd\vec{x}\cdot \dd\vec{x} = \pm \alpha^2 (\dd\zeta)^2 + h_{\sigma\sigmap} \dd{\varrho}_\perp^\sigma \dd{\varrho}_\perp^{\sigmap},
\end{gather}
so $h_{\sigma\sigmap}$ serves as the transverse metric. Below, we assume $\vec{\beta} = 0$, so $\varrho_\perp^\sigma$ and $\varrho^\sigma$ do not need to be distinguished and we can define the inner product on $M^{n-1}_\perp$ as follows:
\begin{gather}\label{eq:innerperp1}
\braket{\psi | \phi}^\perp \doteq 
\int \dd^{n-1}\varrho\,\sqrt{\mdet{h}(\zeta, \vec{\varrho})}\,
\psi^*(\zeta, \vec{\varrho}) \phi(\zeta, \vec{\varrho}).
\end{gather}

We also choose transverse coordinates specifically such that $\vec{\varrho} = \text{const}$. In other words, the coordinate $\vec{\varrho}$ of a given point in space is the initial location of the ray that arrives at this point from the initial transverse surface $M^{n-1}_\perp(0)$. Then, the transverse group velocity is zero,
\begin{gather}\label{eq:vsigma}
V^\sigma \equiv \vec{e}^\sigma \cdot \frac{\dd\vec{\vec{x}}}{\dd\tau} = \frac{\dd \varrho^\sigma}{\dd\tau} = 0,
\end{gather}
where we used \Eqs{eq:dx} and \eq{eq:ort}. This simplifies our equations below. A different definition of $\vec{\varrho}$ is assumed in Papers~II and III, leading to straightforward modifications of the equations, which are not discussed below.

\subsubsection{Quasioptical equation}
\label{sec:qoe}

Suppose a wave beam such that its longitudinal scale $L_\lVert$, which is defined via $\envel_{,\zeta} \sim \envel/L_\lVert$, is much larger than its perpendicular scale $L_\perp$, which is defined via $\envel_{,\sigma} \sim \envel/L_\perp$ and may or may not be the beam global width. (We assume the notation $\envel_{,\zeta} \doteq \pd\envel/\pd \zeta$ and $\envel_{,\sigma} \doteq \pd \envel/\pd \varrho^\sigma$.) We introduce two GO parameters,
\begin{gather}\label{eq:epsilons}
\epsilon_\lVert \doteq \lambda/L_\lVert,
\quad
\epsilon_\perp \doteq \lambda/L_\perp,
\quad
\epsilon_\lVert \ll \epsilon_\perp
\end{gather}
[so the original parameter \eq{eq:epsilon} is $\epsilon = \epsilon_\perp$], and we shall neglect terms smaller than $O(\epsilon_\lVert\epsilon_\perp)$ from now on. Also, we require that either $\alpha$ is chosen as independent of $\vec{\varrho}$ or, more generally, the relative variation of $\alpha$ across the wave beam is small enough such that $\alpha_{,\sigma}$ is negligible. We also assume 
\begin{gather}\label{eq:scalDh}
(D_H)_{|\mu} \lesssim O(\epsilon_\lVert), \quad g_{\alpha\beta,\mu} \lesssim O(\epsilon_\lVert),
\end{gather}
where the derivatives can be taken over the longitudinal or transverse coordinates. (Having $g_{\alpha\beta,\mu} \sim 1/L_\lVert$ is indeed typical in typical applications \cite{tex:mypaper2}.)

Under these assumptions, \Eq{eq:dprime} can be approximated as follows [assuming also \Eq{eq:drOm}]:
\begin{gather}
\oper{\mc{D}}_H \envel = 
- i V\envel_{,\zeta} - \frac{i}{2 \sqrt{\mdet{g}}}\,(\sqrt{\mdet{g}}V)_{,\zeta} \envel 
+ \oper{\mc{G}} \envel.
\end{gather}
Here, $V \doteq V^0$, and we used that $V^\mu \envel_{,\mu} = V \envel_{,\zeta}$ due to \Eq{eq:vsigma}. The operator $\oper{\mc{G}}$ is self-adjoint under the inner product \eq{eq:innerperp1} and given by
\begin{gather}
\oper{\mc{G}} \envel \doteq - \frac{1}{2\mdet{h}^{1/4}}\,[\Deffr^{|\sigma\sigmap} (\mdet{h}^{1/4}\envel)_{,\sigmap}]_{,\sigma}.
\label{eq:Gop}
\end{gather}
(We kept only the transverse derivatives in this term because the longitudinal ones are beyond the assumed accuracy.) Similarly, \Eq{eq:ddprime} becomes
\begin{gather}\label{eq:Dpp}
\oper{\mc{D}}_A \envel = \Deffi
- i \mc{V}_A^\sigma \envel_{,\sigma} - \frac{i}{2 \sqrt{\mdet{h}}}(\sqrt{\mdet{h}}\mc{V}_A^\sigma)_{,\sigma} \envel,
\end{gather}
where the higher-order terms are neglected because $\Deffi$ is assumed small [\Eq{eq:ordering}]. Although the last term in \Eq{eq:Dpp} is negligible within the assumed accuracy, it is retained anyway to make $\oper{\mc{D}}_A$ (not to be confused with $i\oper{\mc{D}}_A$) exactly self-adjoint under the inner product \eq{eq:innerperp1}. Then, \Eq{eq:Dtheta} becomes
\begin{gather}\label{eq:qoe0}
iV \envel_{,\zeta} + \frac{i}{2 \sqrt{\mdet{g}}}\,(\sqrt{\mdet{g}}V)_{,\zeta}\envel
-\oper{\mc{G}}\envel = i\oper{\mc{D}}_A\envel,
\end{gather}
and, as a corollary,
\begin{gather}\label{eq:act2}
\frac{1}{\alpha}\,\frac{\dd}{\dd\zeta} \braket{\psi|\alpha V|\psi}^\perp 
 = 2\braket{\psi|\oper{\mc{D}}_A|\psi}^\perp,
\end{gather}
where we used the fact that $\oper{\mc{G}}$ and $\oper{\mc{D}}_A$ are self-adjoint. In particular, for zero $\Deffi$, \Eq{eq:act2} predicts conservation of the wave-action flux through the beam cross section,
\begin{gather}\label{eq:consJ}
\braket{\envel|\alpha V|\envel}^\perp = {\rm const}.
\end{gather}

Equation \eq{eq:act2} indicates that $\oper{\mc{D}}_A$ cannot be much larger than $\pd/\pd \zeta$, so the correct scaling to assume for the dissipation term is $\Deffi \lesssim O(\epsilon_\lVert)$. Then, \Eq{eq:qoe0} leads to the ``quasioptical scaling''
\begin{gather}
\epsilon_\lVert \sim \epsilon_\perp^2.
\end{gather}
Hence, the difference between $\oper{\mc{D}}_A$ and $\Deffi$ in \Eq{eq:Dpp} is $\lesssim O(\epsilon^3)$, so it is beyond the accuracy of our theory, which is $O(\epsilon^2)$. For this reason, we henceforth adopt \cite{foot:deff} 
\begin{gather}\label{eq:DADeffi}
\oper{\mc{D}}_A = \Deffi,
\end{gather}
which will also simplify our notation. [Also, as a reminder, $\pd_\mu g_{\alpha\beta} \lesssim O(\epsilon_\lVert) \sim \epsilon_\perp^2$, and the parameters of the medium vary similarly.] This leads~o
\begin{gather}\label{eq:qoe}
iV \envel_{,\zeta} + \frac{i}{2 \sqrt{\mdet{g}}}\,(\sqrt{\mdet{g}}V)_{,\zeta}\envel
-\oper{\mc{G}}\envel = i\Deffi\envel.
\end{gather}

Equation \eq{eq:qoe} is a general quasioptical equation for a scalar-wave beam in an inhomogeneous medium. Since it is a \textit{parabolic} equation (it contains only the first-order derivative with respect to $\zeta$), \Eq{eq:qoe} is much easier to solve than \Eq{eq:Dtheta} with the original expressions for $\oper{\mc{D}}_H$ and $\oper{\mc{D}}_A$ given by \Eqs{eq:dprime} and \eq{eq:ddprime}.

\subsubsection{Simplified equations}
\label{sec:ortho}

A typical metric of interest for quasioptical modeling has the form $\smash{g_{\mu\nu} = \delta_{\mu\nu} + R_{\mu\nu\sigma}\varrho^\sigma}$ with $\smash{R_{\mu\nu\sigma} = O(L_\lVert^{-1})}$ \cite{tex:mypaper2}. This satisfies the assumed scaling \eq{eq:scalDh} and corresponds to $\ln \mdet{g} = O(L_\perp/L_\lVert)$. Then,
\begin{gather}
(\ln \mdet{g})_{,\zeta} \sim L_\perp/L_\lVert^2 = O(\epsilon_\lVert^{3/2}),
\end{gather}
which is negligible. [In contrast, $(\ln \mdet{g})_{,\sigma} = \smash{O(\epsilon_\lVert)}$.] The derivatives of $\mdet{g}$ in \Eq{eq:Gop} are negligible too. Hence, the metric factors in the quasioptical equation \eq{eq:qoe} can be replaced with unity. (However, the effect of the metric on the dispersion matrix may remain important in this case; see \Sec{sec:EM}.) Then, \Eq{eq:qoe} becomes
\begin{gather}\label{eq:qoe2}
iV \envel_{,\zeta} + \frac{i}{2}\,V_{,\zeta}\envel
+ \frac{1}{2}\,\Big(\Deffr^{|\sigma\sigmap} \envel_{,\sigmap}\Big){}_{,\sigma} = i\Deffi\envel.
\end{gather}
Using the variable transformation $\phi \doteq \sqrt{V} \envel$, this equation can be further simplified as follows:
\begin{gather}\label{eq:qoe3}
i\phi_{,\zeta} + \frac{1}{2}\,(\Sigma^{\sigma\sigmap}\phi_{,\sigmap})_{,\sigma} = i\Upsilon\phi,
\end{gather}
which is just a dissipative Schr\"odinger equation with
\begin{gather}\label{eq:Sigma}
\Sigma^{\sigma\sigmap} \doteq \Deffr^{|\sigma\sigmap}/V,
\quad
\Upsilon \doteq \Deffi/V.
\end{gather}
[In \Eq{eq:qoe3}, we used that $V_{,\sigma}\phi_{,\sigmap}$ is negligible compared to $V \phi_{,\sigmap \sigma} = O(\epsilon_\perp)$.] Also, \Eq{eq:act2} becomes
\begin{gather}
\frac{\dd}{\dd\zeta} \braket{\phi|\phi}^\perp = 2\braket{\phi|\Upsilon|\phi}^\perp.
\end{gather}

Note that $\Sigma$ can also be expressed alternatively as follows. Suppose one finds the group velocity and constructs a ray-based metric (as described in \Sec{sec:metric}) in the vicinity of a given $\vec{x}$. Then, \Eq{eq:drOm} can be considered as an equation for the local $k_0$ as a function of the transverse wave-vector components $k_\sigma$ (and $\vec{x}$) in this prescribed metric; \ie
\begin{gather}
\Deffr\boldparen{\vec{x}, k_\sigma, k_0(\vec{x}, k_\sigma)} = 0.
\end{gather}
By differentiating this with respect to $k_\sigma$, we obtain $\smash{k_0{}^{|\sigma} = - \smash{\Deffr^{|\sigma}/\Deffr^{|\zeta}}}$. By differentiating the latter formula once again, now with respect to $k_{\sigmap}$, we also obtain
\begin{multline}\label{eq:ddk}
k_0{}^{|\sigma\sigmap} = - \frac{1}{\Deffr^{|\zeta}}\,\Big(
\Deffr^{|\zeta\zeta}k_0{}^{|\sigma}k_0{}^{|\sigmap}
+ \Deffr^{|\zeta\sigmap}k_0{}^{|\sigma}
\\+ \Deffr^{|\zeta\sigma}k_0{}^{|\sigmap}
+ \Deffr^{|\sigma\sigmap}
\Big).
\end{multline}
Recall that $\smash{\Deffr^{|\sigma} = V^\sigma}$ and $V^\sigma = 0$ [\Eq{eq:vsigma}]. Then, $\smash{k_0{}^{|\sigma} = 0}$, and by comparing \Eq{eq:ddk} with \Eq{eq:Sigma} and using \Eq{eq:VP}, one finds that $\smash{k_0{}^{|\sigma\sigmap} = -\Sigma^{\sigma\sigmap}}$. Thus, \Eq{eq:qoe3} can be rewritten as
\begin{gather}\label{eq:schrgen}
 i\phi_{,\zeta} - \frac{1}{2}\,(k_0{}^{|\sigma\sigmap} \phi_{,\sigmap})_{,\sigma} 
 = i\Upsilon \phi.
\end{gather}
In the case of a spacetime problem, $\smash{k_0{}^{|\sigma\sigmap}}$ can also be expressed through the wave frequency in the laboratory frame. Then, the familiar Schr\"odinger equation for the wave envelope \cite{foot:karpman} can be reproduced. We do not discuss it further for this subject is not essential to our paper.

\section{Vector waves}
\label{sec:vector}

\subsection{Hilbert space for vector waves}
\label{sec:vbasic}

Now, let us generalize the above results to vector waves. Suppose an $m$-component wave field $\field \equiv \field(\vec{x})$,~so
\begin{gather}\label{eq:PsiVector}
\field = 
\left(
\begin{array}{c}
\field^1\\
\field^2\\
...\\
\field^m
\end{array}
\right), 
\quad
\ket{\field} =
\left(
\begin{array}{c}
\ket{\field^1}\\
\ket{\field^2}\\
...\\
\ket{\field^m}
\end{array}
\right).
\end{gather}
Here, $\ket{\field}$ is a vector on some Hilbert space $\hilb{H}_m^n$ for which we assume the following general inner product:
\begin{gather}\label{eq:innerm}
\braket{\Psi|\Phi}_m \doteq \int \dd^nx \sqrt{\mdet{g}(\vec{x})}\,\vmetr_{ab}(\vec{x}) \Psi^{a*}(\vec{x}) \Phi^b(\vec{x}).
\end{gather}
Here, the Latin indices that characterize the field components span from 1 to $m$. (This is in contrast with the Greek indices introduced earlier, which characterize the components of $\vec{x}$ and span from 0 to $n-1$.) The matrix $\vmetr(\vec{x})$ can be any $m \times m$ symmetric matrix. It can be considered as an additional metric. This metric does not have to be the same as $g$, as seen already from the fact that $m$ and $n$ do not have to be the same. (For example, \Ref{my:qdiel} describes a six-dimensional field on a three-dimensional spacetime; also see \Sec{sec:noncov}.) This means that the space to which $\field(\vec{x})$ belongs is not necessarily the space tangent to $M^n$. That said, having $\vmetr = g$ is possible as a special case; see Secs.~\ref{sec:covar} and~\ref{sec:statw}. 

Using $\vmetr$ as a metric, we introduce the standard rules for manipulating the Latin indices,
\begin{gather}\label{eq:PsiGamma}
\Psi_a \doteq \vmetr_{ab}\Psi^b, \quad \Psi^a = \vmetr^{ab}\Psi_b,
\end{gather}
where $\vmetr^{ab}$ are elements of $\vmetr^{-1}$. In particular, in the $\vec{x}$ representation, one has
\begin{gather}\label{eq:PsiDag}
\Psi^\dag(\vec{x}) \doteq \braket{\Psi|\vec{x}} = (\Psi_1^*, \Psi_2^*,\, \ldots, \Psi_m^*),
\end{gather}
and we shall also use the following local dot product:
\begin{gather}\label{eq:dotpr}
\Psi \cdot \Phi \doteq \Psi^\dag(\vec{x}) \Phi(\vec{x}) = \Psi_a^*(\vec{x}) \Phi^a(\vec{x}).
\end{gather}
Under this dot product, a matrix $A$ with mixed-index elements $A^a{}_b$ is self-adjoint [$(A\Psi) \cdot \Phi = \Psi \cdot (A\Phi)$] if the matrix with the corresponding lower-index elements, $A_{ab}$, is Hermitian. The difference between Hermitian and self-adjoint matrices can be ignored if the metric $\vmetr$ is Euclidean or pseudo-Euclidean.

\subsection{Operators on $\boldsymbol{\hilb{H}_m^n}$}
\label{sec:oper}

\new{Any operator $\oper{A}$ on $\hilb{H}_m^n$ can be understood as an $m \times m$ matrix with elements $\oper{A}^a{}_b$ which are operators on $\hilb{H}_1^n$. For any given $\oper{A}^a{}_b$, we introduce $\underline{\oper{A}}_{ab} \doteq \vmetr_{ac} \oper{A}^c{}_b$, which is also an operator on $\hilb{H}_1^n$. Correspondingly, two types of Weyl images can be defined, namely, $\weyl_{z}[\oper{A}^a{}_b]$ and $\weyl_{z}[\underline{\oper{A}}_{a b}]$. Below, we assume the notation
\begin{gather}
\underline{A}_{ab} \doteq \weyl_{z}[\underline{\oper{A}}_{ab}],
\quad
A^a{}_b \doteq \gamma^{ac} \underline{A}_{cb}.
\end{gather}
Also, $A$ will be an index-free notation of the matrix with elements $A^a{}_b$, and $\underline{A}$ will be an index-free notation of the matrix with elements $\underline{A}_{ab}$. Importantly, $A^a{}_b$ should not be confused with $\weyl_{z}[\oper{A}^a{}_b]$, because the Weyl symbols do not transform as tensors,
\begin{gather}
\weyl_{z}[\oper{A}^a{}_b] \ne \gamma^{ac} \weyl_{z}[\underline{\oper{A}}_{c b}],
\end{gather}
except for those which are independent of $\vec{p}$. Symbols $\weyl_{z}[\oper{A}^a{}_b]$ will not be used below, so no special notation is introduced for them.
}

Note that by \Eq{eq:innerm}, the following equality holds for any $\Phi$, $\Psi$, and $\oper{A}$ on $\mathcal{H}_m^n$:
\begin{align}
\braket{\Psi |\oper{A}\Phi}_m 
& = \int \dd^nx \sqrt{\mdet{g}}\, \vmetr_{a b} \Psi^{a*} (\oper{A}^b{}_c\Phi^c) \notag \\ 
& = \int \dd^nx \sqrt{\mdet{g}}\, \Psi^{a*} (\underline{\oper{A}}_{a b}\Phi^b) \notag \\
& = \braket{\Psi^a|\underline{\oper{A}}_{a b}\Phi^b},
\end{align}
where we invoked the definition of the inner product on $\hilb{H}_1^n$ [\Eq{eq:inner1}]. By definition of the adjoint operator $\oper{A}^\dag$,
\begin{align}
\braket{\Psi|\oper{A}\Phi}_m
& =\braket{\oper{A}^\dag \Psi|\Phi}_m \notag\\
& =\int \dd^nx\sqrt{\mdet{g}}\,\vmetr_{c b} [(\oper{A}^\dag)^c{}_a\Psi^a]^*\Phi^b\notag\\
& =\int \dd^nx\sqrt{\mdet{g}}\,[(\underline{\oper{A}^\dag})_{b a}\Psi^a]^*\Phi^b.\label{eq:adjoper}
\end{align}
In order to express $\oper{A}^\dag $ through $\oper{A}$, let us represent $\underline{\oper{A}}_{a b}$ in terms of the corresponding Weyl image $\underline{A}_{a b}$ and the Wigner operator $\oper{\Delta}_z$ on $\hilb{H}_1^n$, with $z \equiv (\vec{x}, \vec{p})$ (\Sec{sec:wwt}). Then, \Eq{eq:adjoper} leads~to
\begin{multline}
\braket{\Psi|\oper{A}\Phi}_m
= \int \frac{\dd^{2n} z'}{(2\pi)^n}\,\underline{A}_{a b}(z') \braket{\Psi^a|\oper{\Delta}_{z'}\Phi^b} \\
= \int \frac{\dd^{2n} z'}{(2\pi)^n}\,\braket{\underline{A}_{a b}^*(z')\oper{\Delta}_{z'}\Psi^a|\Phi^b},
\end{multline}
where we used that $\oper{\Delta}_{z'}$ is self-adjoint on $\hilb{H}_1^n$. By comparing this with \Eq{eq:adjoper}, one finds that, on one hand,
\begin{gather}
(\underline{\oper{A}^\dag}){}_{ba} = \int \frac{\dd^{2n} z'}{(2\pi)^n}\,\underline{A}_{a b}^*(z') \oper{\Delta}_{z'}.
\end{gather}
On the other hand, by \Eq{eq:invweyl} for the inverse Wigner--Weyl transform, one has
\begin{gather}
(\underline{\oper{A}^\dag}){}_{ba} = \int \frac{\dd^{2n} z'}{(2\pi)^n}\,\weyl_{z'}[(\underline{\oper{A}^\dag}){}_{ba}]\oper{\Delta}_{z'}.
\end{gather}
Hence, the Weyl image of $\oper{A}^\dag $ is simply the conjugate transpose of the Weyl image of $\oper{A}$ in the sense that
\begin{gather}\label{eq:whwh}
\weyl_z [(\underline{\oper{A}^\dag}){}_{ab}] = \underline{A}_{ba}^*(z).
\end{gather}
In other words, \textit{for operators with both indices lowered}, the operations $\weyl_z$ and $^\dag$ commute. 

Equation \eq{eq:whwh} implies that, if $\oper{A}$ is self-adjoint on $\mc{H}_m^n$, then $\underline{A}$ is a Hermitian matrix, and thus the corresponding matrix $A$ is self-adjoint (\Sec{sec:vbasic}). In particular, this means that the Hermitian and anti-Hermitian parts [\Eq{eq:aux3003}] of an operator are determined by, respectively, the Hermitian and anti-Hermitian parts of its lower-index Weyl symbol. We summarize this as follows:
\begin{gather}
(A_H)^a{}_b = \frac{1}{2}\,\vmetr^{ac}(\underline{A}_{cb} + \underline{A}_{bc}^*),\\
(A_A)^a{}_b = \frac{1}{2i}\,\vmetr^{ac}(\underline{A}_{cb} - \underline{A}_{bc}^*).
\end{gather}
For lower-index matrices, the Hermitian and anti-Hermitian parts are defined as usual:
\begin{gather}
(\underline{A}_H)_{ab}= \frac{1}{2}\,(\underline{A}_{ab} + \underline{A}_{ba}^*),\\
(\underline{A}_A)_{ab} = \frac{1}{2i}\,(\underline{A}_{ab} - \underline{A}_{ba}^*).
\end{gather}

\subsection{Weyl expansion of the dispersion operator}
\label{sec:vweyl}

Now, let us consider the dispersion operator $\oper{D}$ in particular. According to the above definition, it is viewed as an $m \times m$ matrix with elements $\oper{D}^a{}_{b}$. By lowering the index in the envelope equation \eq{eq:weq}, one can write this equation as follows:
\begin{gather}
\underline{\oper{D}}_{ab}\field^b = 0, \quad a = 1, 2, \ldots m,
\end{gather}
where $\underline{\oper{D}}_{ab} \doteq \vmetr_{ac}\oper{D}^c{}_{b}$. Equivalently, this can be written as the envelope equation
\begin{gather}\label{eq:aux3001}
\underline{\oper{\mc{D}}}_{ab}\envel^b = 0, \quad 
\underline{\oper{\mc{D}}}_{ab} \doteq e^{-i\theta}\underline{\oper{D}}_{ab}e^{i\theta}.
\end{gather}
Then, the approximate operators $\underline{\oper{\mc{D}}}_{ab}$ can be expressed just like $\oper{\mc{D}}$ for scalar waves (\Sec{sec:Dscalar}),
\begin{gather}\label{eq:mcdscalm}
\oper{\underline{\mc{D}}}_{ab} \approx \underline{D}_{ab}
+ \frac{1}{2}\left[\oper{p}_\mu (\underline{\mc{V}}^\mu)_{ab}
+(\underline{\mc{V}}^\mu)_{ab} \oper{p}_\mu\right]
+ \frac{1}{2}\,\oper{p}_{\mu}(\underline{\Theta}^{\mu \nu})_{ab}\oper{p}_{\nu}.
\end{gather}
Here, the coefficients are $\vec{x}$-dependent matrices; specifically, $\underline{D}_{ab}(\vec{x})
\doteq \underline{D}_{ab}\boldparen{\vec{x},\vec{k}(\vec{x})}$, and
\begin{gather}
[\underline{\mc{V}}^{\mu}(\vec{x})]_{ab}
\doteq \frac{\pd \underline{D}_{ab}\boldparen{\vec{x},\vec{k}(\vec{x})}}{\pd k_{\mu}},\\
[\underline{\Theta}^{\mu \nu}(\vec{x})]_{ab} 
\doteq \frac{\pd^2 \underline{D}_{ab}\boldparen{\vec{x},\vec{k}(\vec{x})}}{\pd k_{\mu}\pd k_{\nu}}.
\end{gather}
One can also multiply \Eq{eq:aux3001} by $\vmetr^{-1}$ to raise the first index. Then, one obtains
\begin{gather}\label{eq:enveqv}
\oper{\mc{D}}^a{}_b\envel^b = 0, \quad \oper{\mc{D}}^a{}_b \doteq \vmetr^{ac}\oper{\underline{\mc{D}}}_{cb}.
\end{gather}
Note that if $\oper{D}$ is self-adjoint, then $\underline{D}_{ab}$ is Hermitian; then, $\oper{\mc{D}}^a{}_b$ is self-adjoint too. 

\new{Unlike in scalar waves (\Sec{sec:aenv}), $\underline{D}_{ab}(\vec{x}, \vec{k})$ is not a covariant object, \ie in this case, not a true tensor. (For example, see \Sec{sec:EM}.) It is convenient to split it as
\begin{gather}\label{eq:DC}
\underline{D}_{ab}(\vec{x}, \vec{k}) = \underline{\Deff}_{ab}(\vec{x}, \vec{k}) + \underline{C}_{ab}(\vec{x}, \vec{k}),
\end{gather}
where $\underline{\Deff}$ \textit{is} a true tensor, and $\underline{C} = O(\pd\msf{P})$ is a small remainder. [Here $\pd\msf{P}$ means the same as in \Sec{sec:aenv}, so $C = O(\epsilon)$ under the GO ordering, and $C = O(\epsilon_\parallel)$  under the quasioptical ordering.] The splitting \Eq{eq:DC} is not unique and is a matter of convenience; for example, one can define $\underline{\Deff}(\vec{x}, \vec{k})$ simply as the zeroth-order dispersion tensor. Hence, we can redefine
\begin{gather}\label{eq:umcVm}
[\underline{\mc{V}}^{\mu}(\vec{x})]_{ab}
\doteq \frac{\pd \underline{\Deff}_{ab}\boldparen{\vec{x},\vec{k}(\vec{x})}}{\pd k_{\mu}},\\
[\Theta^{\mu \nu}(\vec{x})]_{ab} 
\doteq \frac{\pd^2 \underline{\Deff}_{ab}\boldparen{\vec{x},\vec{k}(\vec{x})}}{\pd k_{\mu}\pd k_{\nu}},
\end{gather}
which is equivalent to the above definitions within the accuracy of our theory. Then, assuming the same orderings as in the scalar-wave case (also see below), the following approximation is enough both for the first-order theory and for the quasioptical theory:
\begin{gather}
\oper{\mc{D}} \approx \Deffr + \oper{\mcu{D}},\\
\oper{\mcu{D}} = C_H + iD_A + \oper{\mcu{D}}_{H1} + \oper{\mcu{D}}_{H2}.\label{eq:mcuD}
\end{gather}
Here, $\Deff \equiv \Deff(\vec{x})$, and $\Deff = \vmetr^{-1} \underline{\Deff}$, so $\Deff^a{}_b(\vec{x}) = \vmetr^{ac}\underline{\Deff}_{c b}(\vec{x})$; similar conventions are assumed for $C$ and $D$. Accordingly, the matrices $\Deffr$ and $D_A$ are self-adjoint, and so are the operators $\oper{\mcu{D}}_{H1}$ and $\oper{\mcu{D}}_{H2}$, which are given~by}
\begin{gather}
\oper{\mcu{D}}_{H1} \doteq \frac{1}{2}\,\vmetr^{-1}(\oper{p}_\mu \underline{\mc{V}}_H^\mu
+ \underline{\mc{V}}_H^\mu \oper{p}_\mu),\\
\oper{\mcu{D}}_{H2} \doteq \frac{1}{2}\,\vmetr^{-1}\oper{p}_{\mu}\underline{\Theta}_H^{\mu \nu} \oper{p}_{\nu}.
\end{gather}
Using $[\vmetr^{-1}, \oper{p}_\mu] = i (\vmetr^{-1})_{,\mu}$ and $(\gamma^{-1}\gamma)_{,\mu} = 0$ to obtain
\begin{gather}\label{eq:elldef}
(\vmetr^{-1})_{,\mu} \vmetr = - \vmetr^{-1} \vmetr_{,\mu} \equiv - \ell_\mu,
\end{gather}
one can also express these as
\begin{gather}
\oper{\mcu{D}}_{H1} = \frac{1}{2}\,
[\oper{p}_\mu \mc{V}_H^\mu(\vec{x}) + \mc{V}_H^\mu(\vec{x}) \oper{p}_\mu 
- i\ell_\mu \mc{V}_H^\mu],\label{eq:DH1}\\
\oper{\mcu{D}}_{H2} \approx \frac{1}{2}\,
\oper{p}_{\mu} \Theta_H^{\mu \nu}(\vec{x}) \oper{p}_{\nu}.\label{eq:DH2}
\end{gather}
(The term $\propto \ell_\mu$ must be kept in $\oper{\mcu{D}}_{H1}$ but a similar term in $\oper{\mcu{D}}_{H2}$ can be neglected.) Also,
\begin{gather}\label{eq:vmugamma}
\mc{V}_H^\mu(\vec{x}) \doteq \vmetr^{-1}\underline{\mc{V}}_H^\mu(\vec{x}),\quad
\Theta_H^{\mu\nu}(\vec{x}) \doteq \vmetr^{-1}\underline{\Theta}_H^{\mu\nu}(\vec{x}),
\end{gather}
and since $\vmetr$ is $\vec{k}$-independent, this leads to
\begin{gather}
[\mc{V}^{\mu}_H(\vec{x})]^a{}_b
\doteq \frac{\pd (\Deffr)^a{}_b\boldparen{\vec{x},\vec{k}(\vec{x})}}{\pd k_{\mu}},\\
[\Theta^{\mu \nu}_H(\vec{x})]^a{}_b 
\doteq \frac{\pd^2 (\Deffr)^a{}_b\boldparen{\vec{x},\vec{k}(\vec{x})}}{\pd k_{\mu}\pd k_{\nu}}.
\end{gather}

If $\oper{\mcu{D}}_{H2}$ is neglected, the envelope equation \eq{eq:enveqv} becomes a Dirac-type equation similar to those considered in the context of XGO \cite{phd:ruiz17, my:covar, my:qdiel, my:qdirac, my:qdirpond} and also, for instance, in \Refs{ref:bridges01, ref:friedland87}. We shall also discuss this limit in \Sec{sec:xgo}.

\subsection{Active and passive modes}

\new{\subsubsection{Basis from the eigenvectors of $\Deffr$}}
\label{sec:diabatic}

Let us assume the GO ordering as in \Eq{eq:ordering}, except that $\Deffr$ and $\Deffi$ are now matrices. Then, $\oper{\mcu{D}} = O(\epsilon)$, so the envelope equation [\Eq{eq:Dtheta}] acquires the form
\begin{gather}\label{eq:enveps}
\Deffr(\vec{x}) \envel = O(\epsilon).
\end{gather}
Hence, it is convenient to express $\envel$ through the eigenvectors of $\Deffr(\vec{x})$. \new{Let us denote these eigenvectors as} $\eta_s(\vec{x})$ and the corresponding eigenvalues as $\Lambda_s(\vec{x})$, so
\begin{gather}
\Deffr \eta_s = \Lambda_s \eta_s.
\end{gather}
[Note that $\eta_s(\vec{x}) = \eta_s\boldparen{\vec{x}, \vec{k}(\vec{x})}$, where $\eta_s(\vec{x}, \vec{p})$ are the eigenvectors of $\Deffr(\vec{x}, \vec{p})$. Likewise, $\Lambda_s(\vec{x}) = \Lambda_s\boldparen{\vec{x}, \vec{k}(\vec{x})}$, where $\Lambda_s(\vec{x}, \vec{p})$ are the eigenvalues of $\Deffr(\vec{x}, \vec{p})$.] Since $\Deffr$ is self-adjoint, it has $m$ eigenvectors $\eta_s$ that form a complete basis $\{\eta_s\}$. Let us also introduce the corresponding dual basis $\{\eta^s\}$ as usual, \ie such that 
\begin{gather}\label{eq:dbasis}
\eta^s \cdot \eta_{s'} = \delta^s_{s'}.
\end{gather}
Then, we can represent $\envel$~as
\begin{gather}\label{eq:psidec}
\envel = \eta_s a^s, \quad a^s = \eta^s \cdot \envel \equiv (\eta^s)_a^* \psi^a,
\end{gather}
where $a^s$ serve as the components of $\envel$ in the basis $\{\eta_s\}$. [Remember that the dot product \eq{eq:dotpr} includes conjugation.] Since $\Deffr$ is self-adjoint, it is always possible to make the basis $\{\eta_s\}$ orthonormal, and this choice is assumed below. Then, $\eta_s \cdot \eta_{s'} = \delta_{ss'}$; thus $\eta_s = \eta^s$, \ie
\begin{gather}
(\eta^s)_a = (\eta_s)_a = \vmetr_{ab}(\eta_s)^b.
\end{gather}
Also, for any two fields $\psi = \eta_s a^s$ and $\phi = \eta_s b^s$, the inner product \eq{eq:innerm} can be written as follows:
\begin{gather}\label{eq:innerdel}
\braket{\psi |\phi}_m 
= \int \dd^nx \sqrt{\mdet{g}(\vec{x})}\,a_s^*(\vec{x}) b^s(\vec{x}),
\end{gather}
where $a_s \doteq \delta_{s s'} a^{s'}$. The matrix $\delta$ with elements $\delta_{ss'}$ serves as a Euclidean metric for manipulating the mode indices $s$ and $s'$. Those should not to be confused with the coordinate indices denoted with Greek letters and also with other Latin indices that are manipulated by the metric $\vmetr$ [\Eq{eq:PsiGamma}].

\subsubsection{Amplitude vectors}
\label{sec:act}

The physical meaning of the expansion coefficients $a^s$ is understood as follows. Consider multiplying \Eq{eq:enveps} by $\eta^s$ from the left. That gives $\Lambda_s a^s = O(\epsilon)$, where no summation over $s$ is assumed. This shows that, for a given $s$, there are two possibilities: either $a^s$ is small or $\Lambda_s$ is small. In the first case, the polarization $\eta_s$ does not correspond to a propagating wave mode \textit{per~se}; the small nonzero projection of $\envel$ on $\eta_s$ is only due to the fact that the wave field is not strictly sinusoidal in inhomogeneous medium. We call such ``modes'' passive. In the second case, the local $\vec{k}(\vec{x})$ approximately satisfies the local dispersion relation $\Lambda_s\boldparen{\vec{x}, \vec{k}(\vec{x})} \approx 0$. Then, $a^s$ can be understood as the local scalar amplitude of an actual GO mode, so that $a^s = O(1)$ is allowed. We call such modes active, and mode conversion occurs when more than one active mode exists. In other words, for a given boundary or initial conditions, active modes are those that are excited resonantly, while passive modes are those that are nonresonant and, thus, adiabatically isolated. (However, this does not mean that the passive-modes amplitudes are simply negligible; see below.) 

Assuming that there are $N \ge 1$ active modes, we shall order them such that they correspond to $s = 1, 2, \ldots, N$ and the remaining $\bar{N} = m - N$ modes with $s = {(N + 1)}, {(N+2)}, \ldots, m$ are passive. Let us also adopt the notation
\begin{gather}
\bar{a}^s = a^{s + N}, \quad \bar{\eta}_s = \eta_{s + N}, \quad \bar{\Lambda}_s = \Lambda_{s + N}
\end{gather}
($s = 1, 2, \ldots, \bar{N}$) for the passive-mode amplitudes, polarizations, and eigenvalues. Then, it is convenient to introduce the following ``amplitude vectors''
\begin{gather}
a \doteq \left(
\begin{array}{c}
a^1\\
\vdots\\[5pt]
a^{N}
\end{array}
\right),
\quad
\bar{a} \doteq \left(
\begin{array}{c}
\bar{a}^1\\
\vdots\\[5pt]
\bar{a}^{\bar{N}}
\end{array}
\right)
\end{gather}
and the corresponding row vectors that are dual to the amplitude vectors under the Euclidean complex dot product; namely,
\begin{gather}\label{eq:adag}
a^\dag = (a_1^*, a_2^*, \ldots, a_N^*), 
\quad
\bar{a}^\dag = (\bar{a}_1^*, \bar{a}_2^*, \ldots, \bar{a}_{\bar{N}}^*),
\\
a_s \doteq \delta_{s s'} a^{s'}, \quad \bar{a}_s \doteq \delta_{s s'} \bar{a}^{s'}.\label{eq:adel}
\end{gather}
Below, we seek to derive an approximate envelope equation in terms of these amplitude vectors.

\subsubsection{Polarization matrices}
\label{sec:xi}

Before we proceed, let us introduce the following notation. First, consider the ``polarization matrices''
\begin{gather}
\Xi = (\eta_1, \eta_2, \ldots, \eta_{N}), 
\quad
\bar{\Xi} = (\bar{\eta}_1, \bar{\eta}_2, \ldots, \bar{\eta}_{\bar{N}}).
\end{gather}
These are non-square matrices that have active- and passive-mode polarizations as their columns; namely, 
\begin{gather}
\Xi^a{}_s = (\eta_s)^a, \quad \bar{\Xi}^a{}_s = (\bar{\eta}_s)^a.
\end{gather}
Using these, \Eq{eq:psidec} can be rewritten compactly as
\begin{gather}\label{eq:psiXia}
\envel = \Xi a + \bar{\Xi}\bar{a}.
\end{gather}
Also consider the auxiliary polarization matrices
\begin{gather}\label{eq:Xiplus}
\Xi^+ \doteq \left(
\begin{array}{c}
\eta^{1*}\\
\vdots\\[5pt]
\eta^{N*}
\end{array}
\right),
\quad
\bar{\Xi}^+ \doteq \left(
\begin{array}{c}
\bar{\eta}^{1*}\\
\vdots\\[5pt]
\bar{\eta}^{\bar{N}*}
\end{array}
\right),
\end{gather}
which have the $\eta^{s*}$ as their rows,
\begin{gather}
(\Xi^+)^s{}_a = (\eta^s)_a^*, \quad (\bar{\Xi}^+)^s{}_a = (\bar{\eta}^s)_a^*.
\end{gather}
Since
\begin{multline}
(\envel^\dag)_a = (\eta_s)_a^* a^{s*} + (\bar{\eta}_s)_a^* \bar{a}^{s*} 
= (\eta^{s})_a^* a_s^* + (\bar{\eta}^{s})_a^* \bar{a}_s^*,
\end{multline}
one obtains
\begin{gather}\label{eq:psidag}
\envel^\dag = a^\dag \Xi^+ + \bar{a}^\dag \bar{\Xi}^+.
\end{gather}
However, note that in general, $^+$ does \textit{not} mean to represent the adjoint in the common sense [cf. \Eq{eq:Xip}].

Next, notice that
\begin{gather}
(\Xi^+ \Xi)^s{}_{s'} = (\eta^s)_a^* (\eta_{s'})^a = \eta^s \cdot \eta_{s'} = \delta^s_{s'},
\end{gather}
where the indices $s$ and $s'$ span from 1 to $N$. Analogous formulas apply to $\bar{\Xi}$. Then,
\begin{gather}
(\bar{\Xi}^+ \Xi)^s{}_{s'} = (\bar{\eta}^s)_a^* (\eta_{s'})^a 
= \bar{\eta}^s \cdot \eta_{s'} = 0,
\end{gather}
and similarly for $\Xi^+ \bar{\Xi}$. In other words, one has
\begin{gather}\label{eq:Xiprop}
\Xi^+ \Xi = \mathbb{1},
\quad
\bar{\Xi}^+ \bar{\Xi} = \mathbb{1},
\quad
\bar{\Xi}^+ \Xi = \mathbb{0},
\quad 
\Xi^+ \bar{\Xi} = \mathbb{0},
\end{gather}
where $\mathbb{1}$ is the unit square matrix and $\mathbb{0}$ is the zero square matrix, correspondingly. [The dimensions of $\mathbb{1}$ and $\mathbb{0}$ are different in different equations.] Hence, if one multiplies \Eq{eq:psiXia} by $\Xi^+$ and, separately, by $\bar{\Xi}^+$, one obtains concise formulas for the amplitude vectors $a$ and $\bar{a}$,
\begin{gather}
a = \Xi^+ \envel, \quad \bar{a} = \bar{\Xi}^+ \envel.
\end{gather}

Another property that we shall use later on is
\begin{gather}
(\Xi^+)^s{}_a 
= \delta^{ss'} (\eta_{s'})^*_a 
= \delta^{ss'} (\eta_{s'})^{b*} \vmetr_{a b}
= (\delta^{-1}\Xi^{\sf H}\vmetr)^s{}_a,\notag
\end{gather}
and similarly for $\bar{\Xi}^+$. Here, $^{\sf H}$ denotes the conjugate transpose ($\Xi^{\sf H} \doteq \Xi^{\intercal*}$), and $\delta^{-1}$ is the Kronecker matrix with upper-index elements $\smash{\delta^{ss'}}$. Hence,
\begin{gather}
\Xi^+ = \delta^{-1}\Xi^{\sf H}\vmetr,
\quad
\bar{\Xi}^+ = \delta^{-1}\bar{\Xi}^{\sf H}\vmetr.\label{eq:Xip}
\end{gather}
The factor $\delta^{-1}$ can be omitted if one ignores the difference between upper and lower indices $s$ and $s'$ that refer to the mode number. If $\vmetr$ is Euclidean, one can also ignore the difference between upper and lower coordinate indices; then, $\Xi^+ = \Xi^{\sf H}$.

As a side remark, note that $\Xi$ and $\bar{\Xi}$ are generally non-square, so $\Pi \doteq \Xi\Xi^+$ and $\bar{\Pi} \doteq \bar{\Xi}\bar{\Xi}^+$ are not unit matrices but rather \textit{projectors}. Indeed, due to \Eqs{eq:Xiprop}, one has $\Pi^2 = \Pi$ and $\bar{\Pi}^2 = \bar{\Pi}$. Also, by applying $\Pi$ and $\bar{\Pi}$ to \Eq{eq:psiXia}, one obtains
\begin{gather}\label{eq:Pis}
\Xi a = \Pi \envel, \quad \bar{\Xi} a = \bar{\Pi} \envel.
\end{gather}
Thus, $\Pi \envel$ is the projection of $\envel$ on the active-mode space, and $\bar{\Pi}\envel$ is the projection of $\envel$ on the passive-mode space.

\subsubsection{Equation for the active modes}
\label{sec:actam}

Using the eigendecomposition theorem and \Eq{eq:Xiprop}, one obtains
\begin{gather}\label{eq:aux1003}
\Deffr = \eta_s \Lambda_s \eta^s
= \Xi \Lambda \Xi^+ + \bar{\Xi} \bar{\Lambda} \bar{\Xi}^+,
\end{gather}
where $\Lambda$ and $\bar{\Lambda}$ are the diagonal eigenvalue matrices, 
\begin{gather}
\Lambda \doteq \text{diag}\,\{\Lambda_1, \Lambda_2, \ldots, \Lambda_{N} \} = O(\epsilon),
\\
\bar{\Lambda} \doteq \text{diag}\,\{\bar{\Lambda}_1, \bar{\Lambda}_2, \ldots, \bar{\Lambda}_{\bar{N}} \} = O(1).
\end{gather}
These and \Eq{eq:Xiprop} also lead to
\begin{gather}\label{eq:DXi}
\Deffr\Xi = \Xi \Lambda = O(\epsilon),\quad
\Deffr\bar{\Xi} = \bar{\Xi} \bar{\Lambda} = O(1).
\end{gather}
Hence, the envelope equation \eq{eq:Dtheta} can be written as
\begin{multline}
0 = \Deffr \Xi a + \Deffr \bar{\Xi}\bar{a} + \oper{\mcu{D}}\Xi a + \oper{\mcu{D}} \bar{\Xi}\bar{a}
\\ = \Xi \Lambda a + \bar{\Xi}\bar{\Lambda}\bar{a} + \oper{\mcu{D}}\Xi a + \oper{\mcu{D}} \bar{\Xi}\bar{a},
\end{multline}
where we used \Eq{eq:DXi}. Let us multiply this by $\Xi^+$ and, separately, by $\bar{\Xi}^+$. Then, due to \Eq{eq:Xiprop}, one obtains
\begin{gather}
\Lambda a + \Xi^+ \oper{\mcu{D}} \Xi a + \Xi^+ \oper{\mcu{D}} \bar{\Xi}\bar{a} = 0,\label{eq:aeq1}\\
\bar{\Lambda}\bar{a} + \bar{\Xi}^+ \oper{\mcu{D}}\Xi a + \bar{\Xi}^+ \oper{\mcu{D}} \bar{\Xi}\bar{a} \label{eq:aeq2}= 0.
\end{gather}
Let us also use \Eq{eq:aeq2} to express $\bar{a}$ through $a$ and substitute the result into \Eq{eq:aeq1}. Since $\bar{a}$ and $\oper{\mcu{D}}$ are both of order $\epsilon$, it is sufficient to solve \Eq{eq:aeq2} for $\bar{a}$ approximately to the leading order,
\begin{gather}
\bar{a} \approx - \bar{\Lambda}^{-1} \bar{\Xi}^+ \oper{\mcu{D}}\Xi a.
\end{gather}
Then, \Eq{eq:aeq1} becomes an equation for just the $N$-dimensional active-mode amplitude vector,
\begin{gather}\label{eq:aux202}
(\Lambda + \Xi^+ \oper{\mcu{D}} \Xi - \Xi^+ \oper{\mcu{D}} \bar{\Xi}\bar{\Lambda}^{-1} \bar{\Xi}^+ \oper{\mcu{D}}\Xi)a = 0.
\end{gather}
As a reminder, this equation is valid up to $O(\epsilon^2)$.

\subsection{Leading-order approximation:\\extended geometrical optics}
\label{sec:xgo}

Upon substituting \Eqs{eq:mcuD} into \Eq{eq:aux202}, we obtain the following equation to lowest (first) order in $\epsilon$:
\begin{gather}\label{eq:aux203}
(\Lambda + i\tLambda + \oper{K})a = 0,
\end{gather}
where we introduced
\begin{gather}
\tLambda \doteq \Xi^+ D_A \Xi,\label{eq:tL}
\\
\oper{K} \doteq \Xi^+ (\oper{\mcu{D}}_{H1} + C_H)\Xi.\label{eq:K}
\end{gather}
As shown in \App{app:aux1}, 
\begin{gather}
\oper{K} a =-i V^{\mu}a_{,\mu}-\frac{i}{2}\,V^{\mu}{}_{;\mu}a - U a,\\
V^\mu \doteq \Xi^+ \mc{V}_H^{\mu} \Xi \approx \Lambda^{|\mu},\label{eq:Vmu}\\
U = \delta^{-1}(\Xi^{\sf H}_{,\mu}\underline{\mc{V}}_H^{\mu}\Xi)_A 
- \Xi^+ C_H\Xi,\label{eq:U}
\end{gather}
where $\Xi$ is considered as a function of $\vec{x}$. [As a reminder, $\Xi^{\sf H} \doteq \Xi^{\intercal*}$ is the conjugate transpose of $\Xi$; $\underline{\mc{V}}_H^{\mu}$ is given by \Eq{eq:umcVm}; and $\delta^{-1}$ is a unit matrix that only raises the mode index.] Alternatively, $\Xi$ can be considered as a function of $(\vec{x}, \vec{k})$. Then, as shown in \App{app:aux2},
\begin{multline}\label{eq:U2}
U \approx \delta^{-1}(
\Lambda_{|\mu} \Xi^{\sf H} \vmetr \Xi^{|\mu}
-\Lambda^{|\mu} \Xi^{\sf H} \vmetr \Xi_{|\mu}
+ \Xi^{{\sf H}|\mu} \underline{\Deff}_H \Xi_{|\mu}
)_A 
\\ - \Xi^+ C_H\Xi,
\end{multline}
where the partial derivatives $_{|\mu}$ and $^{|\mu}$ are defined in \Sec{sec:aenv}. The term $U$ is the Stern--Gerlach Hamiltonian mentioned in the introduction (\Sec{sec:intro}), except here it is generalized to an arbitrary metric. This term causes polarization-driven bending of the ray trajectories, which is missed in traditional GO; also, it causes mode conversion, if more than one active mode is present \cite{phd:ruiz17, my:covar, my:qdiel, my:qdirac, my:qdirpond}. These effects are is discussed in further detail in \Sec{sec:disc}.

More explicitly, \Eq{eq:aux203} can be written as
\begin{gather}\label{eq:aux204}
\Lambda a  - i V^{\mu}a_{,\mu} - \frac{i}{2}\,V^{\mu}{}_{;\mu}a = (U  - i\tLambda) a.
\end{gather}
\new{This generalizes the XGO equation derived in \Refs{phd:ruiz17, my:covar, my:qdiel, my:qdirpond, my:qdirac, my:xo} to dissipative waves and curved coordinates. Since the vector $a$ belongs to the space where the metric is Euclidean by definition (\Sec{sec:act}), the elements of this vector are covariant, \ie invariant with respect to coordinate transformations in the physical space. Same applies to the eigenvalues comprising $\Lambda$; thus, the whole left-hand-side of \Eq{eq:aux204} is covariant too. This means that the operator on the right-hand side is also covariant, which includes its Hermitian and anti-Hermitian parts separately. Hence, $U$ and $\tLambda$ are covariant.}

Also note that \Eq{eq:aux204} is similar to \Eq{eq:gopsi} for scalar waves and has a similar corollary,
\begin{gather}\label{eq:vaflux}
\mc{J}^\mu{}_{;\mu} = -2 a^\dag \tLambda a,
\quad
\mc{J}^\mu \doteq - a^\dag V^\mu a.
\end{gather}
Using \Eqs{eq:PsiDag}, \eq{eq:vmugamma}, \eq{eq:psiXia}, \eq{eq:psidag}, \eq{eq:Pis}, and \eq{eq:Vmu} one obtains
\begin{multline}\label{eq:Jmu}
-\mc{J}^\mu = a^\dag \Xi^+ \mc{V}_H^\mu \Xi a
= (\Pi\envel)^\dag \mc{V}_H^\mu (\Pi\envel)
\\ 
\approx \envel^\dag \mc{V}_H^\mu \envel
= \envel^* \underline{\mc{V}}_H^\mu \envel,
\end{multline}
so \Eq{eq:vaflux} can be viewed as a generalization of \Eq{eq:actgo}. At zero $\tLambda$, the dynamics is conservative, and \Eq{eq:vaflux} reflects conservation of the wave action, or quanta \cite{my:amc}. Accordingly, $\tLambda$ determines the dissipation rate, $\mc{J}^\mu$ can be identified (up to a constant factor) as the action flux density summed over all active modes (for example, see \Sec{sec:noncov}), and the elements of the diagonal matrix $\Lambda^{|\mu}$ are proportional to the group velocities of the corresponding active modes. We shall call them the group velocities (without ``proportional to'') for brevity. We also emphasize that this model applies even when the group velocities of the active modes are very different, unlike the quasioptical model discussed in \Sec{sec:qoa}.

For scalar waves studied in \Sec{sec:weylexp}, we had $\Lambda = \Deffr$ and we defined $\vec{k}(\vec{x})$ such that this term be zero [\Eq{eq:drOm}]. Now, $\Lambda$ is a diagonal \textit{matrix} with $N \ge 1$ nonzero elements, so it cannot be zeroed entirely by imposing just one scalar constraint on $\vec{k}(\vec{x})$. Hence, there can be more than one natural way to define $\vec{k}(\vec{x})$. One aesthetically pleasing and convenient \cite{my:xo} option is to require that $\Lambda$ or $\Lambda - U$ be \textit{traceless}. This amounts to choosing the reference-ray Hamiltonian in \Sec{sec:env} as ${H = \text{tr}\,\Lambda/N}$ or $H = {\text{tr}\,(\Lambda - U)/N}$, correspondingly. [Arbitrary constant factors can be introduced instead of $N$, for that only redefines $\tau$ in \Eq{eq:rtp}.] Another option is to adopt $\Lambda_s = 0$ for some single ${s \le N}$, since all active modes have close wave vectors anyway; then $H = \Lambda_s$. As mentioned in \Sec{sec:env}, all such choices of $\vec{k}(\vec{x})$ are equally justified. Although they lead to slightly different envelope equations, those equations are equivalent in the sense that they all describe the same total field by construction. (One might call this a gauge freedom.) However, for the case $N = 1$, choosing $H = \Lambda - U$ is preferable, as discussed in \Sec{sec:single}.

\subsection{Quasioptical model}
\label{sec:qoa}

\subsubsection{Quasioptical equation}

Suppose that a wave propagates largely as a single beam. This implies that all active modes have group velocities close to their average group velocity $\vec{V}_{\rm avr}$, which can be defined via $N V_{\rm avr}^\mu \doteq \text{tr}\,\Lambda^{|\mu}$. Then,
\begin{gather}
V^\mu = V_{\rm avr}^\mu \mathbb{1} + \Delta V^\mu, \quad \Delta V^\mu \doteq V^\mu - V_{\rm avr}^\mu \mathbb{1},
\end{gather}
where $\Delta V^\mu$ are matrices with eigenvalues much smaller than $V_{\rm avr}$. Let us assume a ray-based coordinate system aligned with $\vec{V}_{\rm avr}$ with the inner product on the transverse space $M^{n-1}_\perp$ defined similarly to \Eq{eq:innerperp1} [cf. also \Eq{eq:innerdel}], namely,
\begin{gather}\label{eq:innerperp2}
\braket{a | b}_N^\perp \doteq \int \dd^{n-1}\varrho\,\sqrt{\mdet{h}(\zeta, \vec{\varrho})}\,
a^*_s(\zeta, \vec{\varrho}) b^s(\zeta, \vec{\varrho}).
\end{gather}
Let us also adopt the quasioptical ordering as we did for scalar waves in \Sec{sec:qo}. In particular, we allow $\Delta V^\mu = O(\epsilon_\perp)$. Then, \Eq{eq:aux202} becomes
\begin{gather}
(\Lambda + i\tLambda + \oper{K} + \oper{\mc{G}})a = 0.
\end{gather}
Here, $\oper{K}$ is defined as in \Eq{eq:K} and can be approximated as follows:
\begin{gather}
\oper{K} a = - iV a_{,\zeta} -\frac{i}{2\sqrt{\mdet{g}}}\,(\sqrt{\mdet{g}}V)_{,\zeta} a + \Delta\oper{K}a,\\
\Delta\oper{K}a \doteq - U a -i \Delta V^{\sigma}a_{,\sigma} -\frac{i}{2\sqrt{\mdet{h}}}\,(\sqrt{\mdet{h}}\Delta V^\sigma)_{,\sigma} a,
\label{eq:dK}\\[5pt]
\Delta V^\sigma = \Xi^+ \Deffr^{|\sigma} \Xi,\label{eq:dV2}
\end{gather}
where $V \doteq V_{\rm avr}^0$ (we used the fact that $V_{\rm avr}^\sigma = 0$ by definition of the ray-based coordinates), $U = O(\epsilon_\lVert)$ is given by \Eq{eq:U2}, and $\tLambda$ is given by \Eq{eq:tL}. Finally,
\begin{gather}
\oper{\mc{G}} \doteq \Xi^+ \oper{\mcu{D}}_{H2} \Xi
- \Xi^+ \oper{\mcu{D}}_{H1} \bar{\Xi}\bar{\Lambda}^{-1} \bar{\Xi}^+ \oper{\mcu{D}}_{H1}\Xi.\label{eq:mcG}
\end{gather}
As shown in \App{app:aux3}, $\oper{\mc{G}}$ can also be simplified as
\begin{gather}
\oper{\mc{G}} a \approx - \frac{1}{2\mdet{h}^{1/4}}\,[\Lambda^{|\sigma\sigmap} (\mdet{h}^{1/4}a)_{,\sigmap}]_{,\sigma},\label{eq:mcGvec}
\end{gather}
so it is also self-adjoint under the inner product \eq{eq:innerperp2}.

In summary then, the quasioptical equation for vector waves can be written as follows:
\begin{gather}\label{eq:vqe}
\Lambda a + i\tLambda a - iV a_{,\zeta}-\frac{i}{2\sqrt{\mdet{g}}}\,(\sqrt{\mdet{g}}V)_{,\zeta} a + \oper{\mc{G}} a + \Delta\oper{K}a = 0.
\end{gather}
Equation \eq{eq:vqe} is the main result of our paper. \new{Like \Eq{eq:aux204}, this equation is covariant within the accuracy of our theory.} As a reminder, $a$ is generally a vector (${\text{dim}\,a = N \ge 1}$), whose elements are the scalar amplitudes of active modes (\Sec{sec:act}); $\Lambda$ is the diagonal matrix formed by the eigenvalues of $\Deffr$; $\tLambda$ is given by \Eq{eq:tL}; $V = N^{-1}\text{tr}\,\Lambda^{|\zeta}$ is a scalar; $^{|\zeta} \doteq \pd/\pd k_0$; $\oper{\mc{G}}$ is given by \Eq{eq:mcGvec}; and $\Delta\oper{K}$ is given by \Eq{eq:dK}. There, $U$ is given by \Eq{eq:U2}, and $\Delta V$ is given by \Eq{eq:dV2}. The derivatives of $\smash{\mdet{h}}$ could be neglected within the assumed accuracy, but a purist might want to retain them in order to keep $\Delta\oper{K}$ and $\oper{\mc{G}}$ precisely self-adjoint under the inner product \eq{eq:innerperp2}.

Like in \Sec{sec:qoe}, we require the $\alpha$ parameter of the ray-based coordinates to be defined as $\vec{\varrho}$-independent. Then, \Eq{eq:vqe} has the following corollary:
\begin{gather}\label{eq:vflc}
\frac{1}{\alpha}\frac{\dd}{\dd\zeta} \braket{a|\alpha V|a}^\perp_N
 = 2\braket{a|\tLambda|a}^\perp_N,
\end{gather}
which is similar to \Eq{eq:act2}. For the case of zero $\tLambda$, \Eq{eq:vflc} predicts conservation of the wave-action flux through the beam cross section,
\begin{gather}
\braket{\psi|\alpha V|\psi}^\perp_N = {\rm const}.
\end{gather}
Unlike in \Eq{eq:consJ}, this is the flux of all active modes combined, and the fluxes of individual active modes may not be conserved separately.

\subsubsection{Simplified equations}

If the metric is nearly Euclidean (or pseudo-Euclidean) as in \Sec{sec:ortho}, we can drop the metric factors and rewrite \Eq{eq:vqe} as follows:
\begin{multline}
\Lambda a + i\tLambda a - iV a_{,\zeta}-\frac{i}{2}\,V_{,\zeta} a 
- \frac{1}{2}\,(\Lambda^{|\sigma\sigmap} a_{,\sigmap})_{,\sigma} \\
- U a -i \Delta V^{\sigma}a_{,\sigma} -\frac{i}{2}\,\Delta V^\sigma{}_{,\sigma} a = 0.
\end{multline}
Using the variable transformation $\phi \doteq \sqrt{V} a$, this equation can be further simplified as
\begin{gather}\label{eq:dsc}
i\phi_{,\zeta} = \oper{\chi} \phi + i \Upsilon \phi.
\end{gather}
Here, $\oper{\chi}$ is an operator self-adjoint under the inner product \eq{eq:innerperp2} and given by
\begin{gather}\notag
\oper{\chi} \phi \approx Q \phi 
- \frac{1}{2}\,(\Sigma^{\sigma\sigmap}\, \phi_{,\sigmap})_{,\sigma} 
- \frac{\Delta V^{\sigma}}{V}\,i \phi_{,\sigma} - \frac{i}{2}\left(\frac{\Delta V^\sigma}{V}\right)_{\!\!,\sigma} \phi.
\end{gather}
Also, $Q$, $\Sigma$, and $\Upsilon$ are self-adjoint matrices given by
\begin{gather}
Q \doteq (\Lambda-U)/V, 
\quad
\Sigma^{\sigma\sigmap} \doteq \Lambda^{|\sigma\sigmap}/V,
\quad
\Upsilon \doteq \tLambda/V.
\end{gather}
Accordingly, \Eq{eq:vflc} becomes 
\begin{gather}\label{eq:vflc2}
\frac{\dd}{\dd\zeta} \braket{\phi | \phi}^\perp_N = 2\braket{\phi | \Upsilon | \phi}^\perp_N,
\end{gather}
so $\smash{\braket{\phi | \phi}^\perp_N}$ is conserved if $\Upsilon$ is zero.

\subsection{Summary and discussion}
\label{sec:disc}

The quasioptical model proposed above is equally applicable to scalar beams, single-mode vector beams (${N = 1}$), and multi-mode vector beams (${N > 1}$). 

\subsubsection{Scalar beams}

In the case of a scalar beam (\Sec{sec:weylexp}), one has $\Xi = 1$,~so
\begin{gather}
\Lambda = \Deffr, \quad \tLambda = \Deffi, \quad U = 0, \quad \Delta V^\sigma = 0.
\end{gather}
Then, the equations from \Sec{sec:qo} are reproduced. In particular, $\Lambda$ is made zero by the choice of $\vec{k}$ [\Eq{eq:drOm}]. This implies that the ray Hamiltonian is $H = \Lambda$, which leads to the following ray equations:
\begin{gather}\label{eq:re0}
\frac{\dd x^\alpha}{\dd\tau} = \frac{\pd \Lambda}{\pd k_\alpha}, 
\quad
\frac{\dd k_\alpha}{\dd\tau} = -\frac{\pd \Lambda}{\pd x^\alpha}.
\end{gather}

\subsubsection{Single-mode vector beams} 
\label{sec:single}

In the case of a single-mode vector beam ($N = 1$), one has $\Xi = \eta$, where $\eta$ is the polarization vector. Then,
\begin{gather}
\Lambda = \eta^\dag\Deffr\eta, \quad \tLambda = \eta^\dag D_A \eta,
\end{gather}
so they are scalars. Also, $\Delta V^\sigma = 0$, but $U$ is generally a nonzero scalar function. There are two natural ways to define $\vec{k}$ in this case. One is to require that $\Lambda = 0$, \ie to adopt $H = \Lambda$. In this case, $U$ is left in the envelope equation and can intensify the envelope inhomogeneity in the direction perpendicular to the group velocity. This may eventually result in the violation of the envelope approximation. (For multi-mode beams, this is less of a concern because they typically split into single-mode beams before the issue becomes significant.) Hence, it is potentially advantageous to define $\vec{k}$ by requiring $\Lambda - U = 0$, \ie by adopting $H = \Lambda - U$. In this case, the amplitude equation is identical to that of a scalar wave while the effect of $U$ is absorbed in the reference phase, leading to modified ray equations,
\begin{gather}\label{eq:raysU}
\frac{\dd x^\alpha}{\dd\tau} = \frac{\pd (\Lambda - U)}{\pd k_\alpha}, 
\quad
\frac{\dd k_\alpha}{\dd\tau} = -\frac{\pd (\Lambda - U)}{\pd x^\alpha}.
\end{gather}
The effect of $U$ on the ray trajectories is known as the (spin) Hall effect of light in optics \cite{ref:onoda04, ref:bliokh08, ref:bliokh15, ref:gosselin07, ref:duval17, foot:marius}. In plasma physics, this effect is typically neglected. (To our knowledge, all existing ray-tracing codes ignore $U$ entirely.)

Finally, note that the ray dynamics can also be represented in an alternative form. Since $\Lambda$ is a scalar, one can rewrite \Eq{eq:U2} for $U(\vec{x}, \vec{k})$ as follows:
\begin{multline}
U = U_0 - \Lambda^{|\mu} \text{Im}\,(\eta^\dag \eta_{|\mu}) + \Lambda_{|\mu}\text{Im}\,(\eta^\dag \eta^{|\mu})\\
\approx U_0 - \dot{x}^\mu \msf{A}_\mu^{(x)}
- \dot{k}_\mu\msf{A}^\mu_{(k)}.
\end{multline}
Since there is only one mode, there is no mode index to manipulate, so the ``metric'' factor $\delta$ has been dropped. Likewise, the anti-Hermitian parts are simply the imaginary parts in this case. We also substituted the GO ray equations \eq{eq:re0}, where we replaced $\dd/\dd \tau$ with dots, and adopted
\begin{gather} 
U_0 \doteq \text{Im}\,(\eta^{{\sf H}|\mu} \underline{\Deff}_H \eta_{|\mu}) - \eta^{\sf H}\underline{C}_H\eta, 
\\ 
\msf{A}_\mu^{(x)} \doteq  \text{Im}\,(\eta^\dag \eta_{|\mu}),
\quad
\msf{A}^\mu_{(k)} \doteq \text{Im}\,(\eta^\dag \eta^{|\mu}).
\end{gather}
Then, the phase-space Lagrangian of a ray, $\msf{L}[\vec{x}, \vec{k}] = k_\mu \dot{x}^\mu - H$, has a non-canonical structure, namely,
\begin{gather}\label{eq:aux3333}
\msf{L} =  \big[k_\mu + \msf{A}_\mu^{(x)}\big] \dot{x}^\mu + \msf{A}^\mu_{(k)}\dot{k}_\mu - (\Lambda - U_0).
\end{gather}
The equations originating from a special case of this non-canonical phase-space Lagrangian were used, for example, in \Ref{ref:bliokh08} to study the Hall effect for light propagating in non-birefringent material. A comparison of the resulting non-canonical ray equations with the canonical ray equations \eq{eq:raysU} can be found in \Ref{my:qdiel}. Also notably, the terms $\smash{\msf{A}_\mu^{(x)}}$ and $\smash{\msf{A}^\mu_{(k)}}$ are known as the Berry connection and are linked to the Berry phase, or the geometrical phase of light \cite{ref:bliokh08, my:qdiel, ref:littlejohn91}.

\subsubsection{Multi-mode vector beams} 

In the case of multiple active modes ($N > 1$), $a$ is an $N$-dimensional vector, and the coefficients in the amplitude equation are matrices (except for $V$, which is a scalar). In particular, the matrices $\tLambda$, $\Delta V^\sigma$, and $U$ are generally nondiagonal, so they can cause mode conversion. In the simplest case, when both the transverse gradients and dissipation are negligible, this process is most transparently described by \Eq{eq:dsc}, which then becomes
\begin{gather}\label{eq:sch3}
i\phi_{,\zeta} = Q \phi,
\end{gather}
with $Q$ being a self-adjoint matrix. The modes decouple when $Q$ is close to diagonal. More generally, the dynamics governed by \Eq{eq:sch3} is similar to that of an $N$-level quantum system with a Hamiltonian $Q$ (and to the dynamics of $N$ coupled classical oscillators whose parameters do not change too rapidly with $\zeta$ \cite{my:wkin}). Alternatively, it can be mapped to a precession equation for a real $(N^2 - 1)$-dimensional ``spin'' vector \cite{my:qdirac}. This approach is particularly intuitive at $N = 2$, when the spin is three-dimensional. A detailed analysis of mode conversion for this case can be found in \Ref{my:xo}.

As a reminder, the model presented here (\Sec{sec:qoa}) relies on the assumption that the group velocities of the active modes are close to each other. Otherwise, beam splitting occurs rapidly, and the assumption that $a_{,\zeta} \ll a_{,\sigma}$ does not hold. The quasioptical description is inapplicable to such beams; however, the first-order XGO model described in \Sec{sec:xgo} can be used.

\section{Electromagnetic waves}
\label{sec:EM}

\subsection{Covariant formulation}
\label{sec:covar}

Here, we shall explain how the above theory applies to EM waves. We start with Maxwell's equation \cite{book:landau2}
\begin{gather}
\frac{1}{\sqrt{\mdet{g}}}\,\pd_{\alpha}(
\sqrt{\mdet{g}}F^{\alpha \beta}
) = - \frac{4\pi}{c}\, J^{\beta},
\end{gather}
where $F^{\alpha \beta}$ is the EM tensor, namely,
\begin{gather}
F^{\alpha \beta} = g^{\alpha \mu} g^{\beta \nu} F_{\mu \nu},
\quad
F_{\alpha \beta}\doteq \pd_{\alpha}A_{\beta}-\pd_{\beta}A_{\alpha},
\end{gather}
$A$ is the vector potential, and $J$ is the current density. This equation can also be represented as follows:
\begin{gather}
\oper{D}_{\rm vac}^{(A)}A = -\frac{4\pi}{c}\, J.
\end{gather}
Here $\oper{D}_{\rm vac}^{(A)}$ is the vacuum dispersion operator,
\begin{multline}
[\oper{D}_{\rm vac}^{(A)}]^{\alpha}{}_{\beta} A^{\beta}
=\frac{1}{\sqrt{\mdet{g}}}\,\pd_{\lambda}
\big\{
\sqrt{\mdet{g}}\,g^{\alpha \mu}g^{\lambda \nu}\\
\times[\pd_{\nu}(g_{\mu \beta}A^{\beta})-\pd_{\mu}(g_{\nu \beta}A^{\beta})]
\big\},
\end{multline}
which is self-adjoint under the inner product \eq{eq:innerm}. By assuming $J = c\oper{\vartheta} A$, were $\oper{\vartheta}$ is some linear operator, one can cast the equation for $A$ in the form \eq{eq:weq}, namely,
\begin{gather}\label{eq:AD}
[\oper{D}_{\rm vac}^{(A)} + 4\pi \oper{\vartheta}\,] A = 0.
\end{gather}

\new{Hence, the theory developed in the previous sections readily applies, specifically, with the dimension of the configuration space being $n = 4$, the dimension of the vector field $A$ being $m = 4$ (so $m = n$), and $\vmetr = g$. The corresponding Weyl symbols are calculated as follows. It can be shown straightforwardly that $\smash{[\oper{\underline{D}}_{\rm vac}^{(A)}]{}_{\alpha \beta}\doteq g_{\alpha \gamma}[\oper{D}_{\rm vac}^{(A)}]^{\gamma}{}_{\beta}}$ is approximately given by
\begin{multline}
[\oper{\underline{D}}_{\rm vac}^{(A)}]_{\alpha \beta} \approx 
\oper{p}_{\alpha}\oper{p}_{\beta}-\oper{p}_{\mu}g^{\mu \nu}g_{\alpha \beta}\oper{p}_{\nu} 
\\
+ i(q_{\alpha}\oper{p}_{\beta}-q_{\beta}\oper{p}_{\alpha})
+ i[(\ell_{\beta})^{\mu}{}_{\alpha}-(\ell_{\alpha})^{\mu}{}_{\beta}
]\oper{p}_{\mu}.
\end{multline}
Here, we omitted second-order derivatives of $g$, which are negligible within the accuracy of our theory. Also, $q$ is defined as in \Eq{eq:qdef}, and $\ell_\mu \doteq g^{-1} g_{,\mu}$, which is in agreement with \Eq{eq:elldef} because $g = \vmetr$. Finally, in the form introduced in \Sec{sec:vweyl}, the above result is
\begin{gather}
[\underline{D}^{(A)}]_{\alpha \beta} \approx p_{\alpha}p_{\beta}-g_{\alpha \beta}g^{\mu \nu}p_{\mu}p_{\nu}
+4\pi (\underline{\vartheta}_0)_{\alpha \beta},\notag\\
C^{(A)}_{\alpha\beta} \approx 
i (q_{\alpha}k_{\beta}-q_{\beta}k_{\gamma})
+i [(\ell_{\beta})^{\mu}{}_{\alpha}-(\ell_{\alpha})^{\mu}{}_{\beta}]k_{\mu}
+4\pi (\Delta \underline{\vartheta})_{\alpha \beta},
\notag
\end{gather}
where $(\underline{\vartheta}_0)_{\alpha \beta}$ is the GO limit of $\underline{\vartheta}$, $(\Delta \underline{\vartheta})_{\alpha \beta}$ is the remaining part of $\underline{\vartheta}$, and $\underline{\vartheta}$ itself is the symbol of $\underline{\oper{\vartheta}}$.}

\new{As a side remark, these equations can be used to calculate the Hall effect of light in gravitational fields, \ie the deviation of vacuum light rays from geodesics in curved spacetime. (Special cases of this effect are discussed in \Refs{ref:gosselin07, ref:duval17, foot:marius}.)}

\subsection{Non-covariant formulation}
\label{sec:noncov}

As a special case, suppose a metric of the form
\begin{gather}
g = \left(
\begin{array}{cc}
 -1 & \vec{0} \\
 \vec{0} & \boldsymbol{h} \\
\end{array}
\right)
\end{gather}
with time-independent spatial metric $\boldsymbol{h}$. Then, the above equations can be simplified as follows. Let us assume the Weyl gauge ($A_0 = 0$). Then, it is sufficient to consider just the spatial part of the four-dimensional \Eq{eq:AD} and replace the vector potential with the electric field
\begin{gather}
E^a \doteq F^{0 a} = -i \oper{p}_0 A^a. 
\end{gather}
(As usual, $\oper{p}_0\doteq -i c^{-1}\pd_t$, which is a self-adjoint operator. Assuming we work with wave fields that have zero time average, $\oper{p}_0$ can also be considered reversible.) Unlike in the previous sections, we now use the Latin indices to denote spatial components, and we shall adopt the bold font for spatial vectors and matrices when using index-free notation. Specifically, one obtains
\begin{gather}
\oper{\vec{D}}^{(E)} \vec{E}=0,\\
\oper{\vec{D}}^{(E)}\doteq (\oper{p}_0)^{-1} [\oper{\vec{D}}_{\rm vac}^{(A)} + 4\pi \oper{\vec{\vartheta}}\,](\oper{p}_0)^{-1},
\end{gather}
where we also multiplied the equation by $(\oper{p}_0)^{-1}$. This has the form \eq{eq:weq} with $n = 4$, $m = 3$, and $\vmetr = \boldsymbol{h}$.

Let us introduce the conductivity operator $\oper{\vec{\sigma}}$ via $\vec{J} = \oper{\vec{\sigma}}\vec{E}$ and notice that $4 \pi \oper{\vec{\sigma}} = i c \oper{p}_0\oper{\vec{\chi}}$ by definition of the susceptibility operator $\oper{\vec{\chi}}$. Then, $4\pi \oper{\vec{\vartheta}} = \oper{p}_0\oper{\vec{\chi}}\oper{p}_0$, so $\oper{\vec{D}}^{(E)}$ can be expressed as follows:
\begin{gather}
\oper{\vec{D}}^{(E)} \doteq(\oper{p}_0)^{-1} \oper{\vec{D}}_{\rm vac}^{(A)} (\oper{p}_0)^{-1} + \oper{\vec{\chi}},
\end{gather}
or equivalently, $\oper{\vec{D}}^{(E)} = (\oper{p}_0)^{-2} \oper{\vec{D}}_{\rm vac}^{(A)} + \oper{\vec{\chi}}$. The corresponding Weyl image is $\smash{\underline{D}^{(E)}_{a b} = (p_0){}^{-2} [\underline{D}_{\rm vac}^{(A)}]_{a b} + \underline{\chi}_{a b}}$, or
\begin{gather}\label{eq:aux3005}
\underline{D}^{(E)}_{a b} 
= (p_0){}^{-2} \big[p_a p_b - p_c p_d h^{c d} h_{a b} + \underline{C}^{(A)}_{a b}\big] + \underline{\varepsilon}_{a b},
\end{gather}
\new{where $\smash{\underline{\varepsilon}_{a b} \doteq h_{a b} + (\underline{\chi}_0)_{a b}}$ serves as the dielectric tensor and $\underline{\chi}_0$ is the GO limit of $\underline{\chi}$.}

\new{Using the fact that the dispersion operator is defined only up to a constant factor, let us introduce an additional factor $1/(16\pi)$. Then, \Eq{eq:aux3005} becomes
\begin{gather}
\underline{D}^{(E)}_{a b} = 
\underline{\Deff}^{(E)}_{a b}+\underline{C}^{(E)}_{a b},\\
\underline{\Deff}^{(E)}_{a b} 
= \frac{1}{16\pi p_0^2}\,
(p_a p_b - p_c p_d h^{c d} h_{a b})
+\frac{\underline{\varepsilon}_{a b}}{16\pi},\\
\underline{C}^{(E)}_{a b} \doteq \frac{1}{16\pi p_0^2}\,
\underline{C}^{(A)}_{a b}.
\end{gather}
The quantity $\underline{C}^{(A)}_{a b}$ was derived in \Sec{sec:covar}, and the vector $q$ [\Eq{eq:qdef}] and the matrices $\ell_a$ [\Eq{eq:elldef}] can be written as
\begin{gather}
q_a = \frac{1}{4}\,\pd_a \ln |\det \boldsymbol{h}\,|,
\quad
\ell_a = \boldsymbol{h}^{-1} \pd_a\boldsymbol{h},
\end{gather}
so they both can be of order $\epsilon_\lVert$; hence, $C = O(\epsilon_\lVert)$. Note that such $C$ is generally non-negligible even for near-Euclidean metrics such as those discussed in \Sec{sec:ortho}.}

From \Eq{eq:Jmu}, we obtain 
\begin{gather}\label{eq:JMUA}
\mc{J}^{\alpha} = - E^{a*}E^b\,\frac{\pd}{\pd k_{\alpha}}\, [\underline{D}_H^{(E)}(t, \vec{x}, \omega, \vec{k})]_{a b},
\end{gather}
becomes as the true action-flux density \cite{my:amc}. As a reminder, the action density $\mc{I}\doteq \mc{J}^0$ can be written as
\begin{gather}
\mc{I} \approx 
\frac{E^{a*}E^b}{16\pi \omega^2}\,
\frac{\pd}{\pd \omega}\, \big\{\omega^2 [\underline{\varepsilon}_H(t, \vec{x}, \omega, \vec{k})]_{a b}\big\}
,
\end{gather}
where we used $[\underline{D}_H^{(E)}]_{a b} E^b \approx 0$. Using the latter and Faraday's law $\vec{B} \approx c \vec{k} \times \vec{E}/\omega$ for the magnetic field $\vec{B}$, one can also rewrite $\mc{I}$ in another common form \cite{book:stix},
\begin{gather}
\mc{I} \approx 
\frac{E^{a*}E^b}{16\pi \omega}\,
\frac{\pd}{\pd\omega}\, \big\{\omega [\underline{\varepsilon}_H(t, \vec{x}, \omega, \vec{k})]_{a b}\big\}
+\frac{ B_a^* B^a}{16\pi \omega}.
\end{gather}
One can also show \cite{my:amc} that the spatial component of the action flux density \eq{eq:JMUA} can be expressed~as
\begin{gather}
\vec{\mc{J}} \approx \frac{\vec{S}}{\omega} 
- \frac{E^{a*}E^b}{16\pi}\,\frac{\pd}{\pd \vec{k}}\, [\underline{\varepsilon}_H(t, \vec{x}, \omega, \vec{k})]_{a b},
\end{gather}
where $\vec{S}$ is the (time- or space-averaged) Poynting vector,
\begin{gather}
\vec{S} = \frac{c}{8\pi}\,\text{Re}\,(\vec{E} \times \vec{B}^*).
\end{gather}
The GO equation \eq{eq:vaflux} can be written as
\begin{gather}\notag
\frac{\pd \mc{I}}{\pd t} + \frac{1}{\sqrt{\mdet{h}}}\,
\frac{\pd}{\pd x^a}
(\sqrt{\mdet{h}} \mc{J}^a)
= -\frac{E^{a*}E^b}{8\pi}\,[\underline{\varepsilon}_A(t, \vec{x}, \omega, \vec{k})]_{a b}.
\end{gather}
The corresponding density of the wave energy is $\omega \mc{I}$, and the density of the energy flux is $\omega \vec{\mc{J}}$, as flows from the variational principle \cite{my:amc}. 

\subsection{Stationary waves}
\label{sec:statw}

For stationary waves with fixed frequency, one has $\smash{\oper{p}_0} = - \omega/c$ and $\omega$ can be treated as a constant parameter. Such waves can be studied on the three-dimensional configuration space, namely, the physical space $\vec{x}$. In this case, $n = m = 3$ and $\gamma = \boldsymbol{h}$. If there is no spatial dispersion, then $\smash{\oper{\vec{\chi}} = \vec{\chi}(\boper{x})}$, where the dependence on $\omega$ is assumed but not emphasized. Assuming that the underlying space is flat (which is always the case in laboratory applications), the symbol $\chi^a{}_b$ in this case is identical to the susceptibility of the homogeneous medium. (Other formulations of the susceptibility have also been proposed in this case, \eg for light in nondispersive dielectric media \cite{my:qdiel} and for waves in cold plasmas \cite{my:covar, phd:ruiz17, ref:friedland88}.)

\new{These properties can also be extended, approximately, to media with \textit{weak} spatial dispersion, \ie such that
\begin{gather}\label{eq:epsdel}
\underline{\chi}_{ab} = \smash{\underline{\chi}^{(0)}_{ab}(\vec{x})} + \underline{\chi}_{ab}^{(k)} (\vec{x}, \vec{p}),
\quad
\underline{\chi}_{ab}^{(k)} \ll \underline{\chi}_{ab}^{(0)}.
\end{gather}
For example, in plasma, $\smash{\underline{\chi}_{ab}^{(k)}}$ is proportional to the temperature, and for EM waves, thermal effects often can be considered as small perturbations \cite{book:stix}. Although the Weyl symbol $\smash{\underline{\chi}_{ab}^{(k)}}$ is, strictly speaking, different from the corresponding part of the susceptibility in homogeneous medium, the difference is of order $\smash{\epsilon_\parallel\underline{\chi}_{ab}^{(k)}}$, so it is much smaller than $\epsilon_\parallel$ and thus can be neglected. In other words, one can simply replace the small $\smash{\underline{\chi}_{ab}^{(k)}}$ with its GO limit, which is usually well known \cite{book:stix}.}

In principle, our general method is also applicable to waves in media with strong spatial dispersion. However, $\smash{\oper{\vec{\chi}}}$ may be hard to calculate in that case unless a medium is homogeneous, and no extrapolation of a known homogeneous-medium model is guaranteed to yield the true general~$\smash{\oper{\vec{\chi}}}$.

\subsection{Waves in flat space}
\label{sec:flat}

Let us also discuss the special case when the configuration space is flat and \Eq{eq:epsdel} applies, as in typical laboratory applications. Although the ray-based metric $\boldsymbol{h}$ is still non-Euclidean in this case (unless the reference rays are straight), one can use the flatness of the configuration space in the following sense. In the Euclidean laboratory coordinates $\tilde{x}$, the spatial metric has the form $\smash{\tilde{h}_{ab} = \delta_{ab}}$ and $\smash{\tilde{C}}$ vanishes. (Here and further, the tildes denote that the corresponding quantities are evaluated in the laboratory coordinates. Note that Papers~II and III assume a different notation; there, tilded are quantities in the ray-based coordinates, and non-tilded are those in the laboratory coordinates.) This simplifies the expression for the dispersion matrix and makes $D_H$ a true tensor (modulo the insignificant corrections caused by weak spatial dispersion discussed in \Sec{sec:statw}). \new{Then, since $U$ and $\tLambda$ are frame-invariant (\Sec{sec:xgo}), one can calculate them in the laboratory coordinates using
\begin{gather}\notag
U = \tilde{U} \approx (
\tilde{\Lambda}_{|\mu} \tilde{\Xi}^{\sf H} \tilde{\Xi}^{|\mu}
-\tilde{\Lambda}^{|\mu} \tilde{\Xi}^{\sf H} \tilde{\Xi}_{|\mu}
+ \tilde{\Xi}^{{\sf H}|\mu} {\tilde{\Deff}}_H \tilde{\Xi}_{|\mu}
)_A,\\
\tLambda = \tilde{\tLambda} = \tilde{\Xi}^{\sf H} \tilde{\Deff}_A \tilde{\Xi},\notag
\end{gather}
where both $\delta^{-1}$ and $\tilde{\vmetr}$ are omitted for simplicity, since they are unit matrices. This also extends to the quasioptical model. Since calculations in the laboratory coordinates are generally easier than the corresponding calculations in the ray-based coordinates, this simplifies applications of our theory to numerical simulations, as discussed further in Paper~II.}

\new{Specific applications of this theory, including those to waves in fusion plasmas, will be discussed in future publications. Here, we only note that our formulation relies on the GO ordering and thus is inapplicable to wave transformations near caustics, for example, as in the O--X conversion caused by the density gradient in dense magnetized plasmas \cite{ref:preinhaelter73}. In contrast, the O--X conversion caused by the magnetic shear in dilute plasmas \cite{my:xo} can be modeled naturally, as also demonstrated explicitly in Paper~III. Any adiabatic transformations of waves along continuous dispersion curves can also be modeled naturally, specifically, within the single-mode approximation summarized in \Sec{sec:single}. (In special cases like the X--B transformation \cite{book:stix}, when the group velocity becomes zero on a ray while $k$ remains large, one may want to replace the coordinate $\zeta$ with $\tau$, where $\dd\tau = \dd\zeta/V$, to remove the singularity in the amplitude equation.) In all these cases, our theory is expected to yield results that agree with full-wave modeling up to local errors of order $O(\epsilon_\parallel)$, because this is the accuracy within which the passive-mode contribution is calculated (\Sec{sec:actam}).}

\section{Conclusions}
\label{sec:conc}

In summary, we propose a quasioptical theory of mode-converting wave beams in inhomogeneous media such as plasma. This includes the following. For any given dispersion operator $\oper{D}$ that governs the original wave field $\field$, we explicitly calculate the approximate operator $\oper{\mc{D}}$ that governs the wave envelope $\envel$ to the second order in the GO parameter $\epsilon$. Then, we further simplify this envelope operator by assuming that the gradient of $\psi$ transverse to the local group velocity is much larger than the corresponding parallel gradient. This leads to a parabolic differential equation for $\envel$ (``quasioptical equation'') in the basis of the GO polarization vectors [\Eq{eq:psidec}]. Our main results can be found in \Sec{sec:qoa}, which includes the general quasioptical equation \eq{eq:vqe} and its special case \eq{eq:dsc}. Scalar and mode-converting vector beams are described on the same footing (\Sec{sec:disc}). We also explain how to apply this model to EM waves considered as a special case (\Sec{sec:EM}). In the follow-up papers, we report successful quasioptical modeling of radiofrequency wave beams in magnetized plasma based on this theory.

\new{\section{Acknowledgments}}

The first author (IYD) acknowledges the support and hospitality of the National Institute of Fusion Science, Japan. The authors are also thankful to M.~A. Oancea for stimulating discussions regarding the XGO generalization to non-Euclidean metrics. The work was supported by JSPS KAKENHI Grant Number JP17H03514. The work was also supported by the U.S. DOE through Contract No. DE-AC02-09CH11466 and by the Laboratory Directed Research and Development program at Sandia National Laboratories. Sandia National Laboratories is a multimission laboratory managed and operated by National Technology and Engineering Solutions of Sandia, LLC., a wholly owned subsidiary of Honeywell International, Inc., for the U.S. DOE National Nuclear Security Administration under contract DE-NA-0003525. This paper describes objective technical results and analysis. Any subjective views or opinions that might be expressed in the paper do not necessarily represent the views of the U.S. Department of Energy or the United States Government.

\mbox{}

\appendix

\section{Summary of selected notations}
\label{app:sumnot}

Here, we present a summary of selected notations used in the main text. 

\subsection{Basic symbols}

\begin{itemize}
\item
$\doteq$ denotes a definition.
\item
$\oper{\phantom{A}}$ denotes an operator.
\item 
$^\dag$ denotes either a dual vector [\Eqs{eq:PsiDag} and \eq{eq:adag}] or an adjoint operator.
\item
$^\intercal$ denotes the matrix transpose.
\item
$^{\sf H} = {}^{\intercal*}$ denotes the conjugate matrix transpose.
\item
$^+$ is used only in $\Xi^+$ and $\bar{\Xi}^+$ defined in \Eqs{eq:Xiplus}.
\item
$_H$ and $_A$ denote the Hermitian and anti-Hermitian parts of an operator [\Eqs{eq:aux3003}] or those of a matrix (\Sec{sec:oper}). When applied to a scalar, $_H$ and $_A$ denote the real and imaginary parts, correspondingly.
\item
The underline notation $\underline{\phantom{A}}$ is explained in \Sec{sec:oper} and is used for matrices with lower indices only.
\item
$\weyl_z$ and $\weyl^{-1}$ denote the Wigner--Weyl transform and its inverse (\Sec{sec:wwt}), respectively.
\end{itemize}

\subsection{Index manipulation}

On the $n$-dimensional configuration space $M^n$, we assume a general metric tensor $g$ with components $g_{\mu\nu}$. For (co)vector fields on the space (co)tangent to $M^n$, the indices are manipulated as usual, namely, via
\begin{gather}\label{eq:aux601}
X_\mu = g_{\mu\nu} X^\nu, \quad \mu,\nu = 0,1, \ldots, (n-1).
\end{gather}
(Summation over repeated indices is always assumed unless specified otherwise.) For vector waves, we also introduce an additional vector space that may or may not be the same as the space tangent to $M^n$ (\Sec{sec:vbasic}). On that space, an additional metric $\vmetr$ is introduced, and the indices are manipulated~as follows:
\begin{gather}\label{eq:aux602}
\Psi_a = \vmetr_{ab}\Psi^b, \quad a,b = 1,2, \ldots, m.
\end{gather}
In special cases, \Eqs{eq:aux601} and \eq{eq:aux602} can be equivalent, but that is not a generic situation. 

We also introduce the Euclidean metric $\delta$ on the space of amplitude vectors $a$ (\Sec{sec:act}),
\begin{gather}
a_s = \delta_{ss'}a^{s'}.
\end{gather}
For active modes, which are of our primary interest, the mode indices $s$ and $s'$ range from 1 to $N$. The distinction between $a^s$ and $a_s$ is made only for aesthetic reasons (consistency of notation) and can be neglected otherwise. 

\subsection{Derivatives}

For spatial derivatives, we use the notation that is standard, for example, in general relativity \cite{book:misner77}. In particular, $X^\mu{}_{;\mu}$ is the divergence; namely,
\begin{gather}
X^\mu{}_{;\mu} = \frac{1}{\sqrt{\mdet{g}}}\, (\sqrt{\mdet{g}} X^\mu)_{,\mu}.
\end{gather}
Here, $\mdet{g} \doteq |\det g\,|$, and
\begin{gather}
f_{,\mu} \equiv \pd_\mu f \doteq \frac{\pd f}{\pd x^\mu},
\quad
f_{,\mu\nu} \doteq \frac{\pd^2 f}{\pd x^\mu \pd x^\nu}.
\end{gather}
For functions of the form $f(\vec{x}, \vec{k})$, we also introduce the following partial derivatives:
\begin{gather}
f_{|\mu} \doteq \frac{\pd f(\vec{x}, \vec{k})}{\pd x^\mu},
\quad
f_{|\mu\nu} \doteq \frac{\pd^2 f(\vec{x}, \vec{k})}{\pd x^\mu\pd x^\nu},\\
f^{|\mu} \doteq \frac{\pd f(\vec{x}, \vec{k})}{\pd k_\mu},
\quad
f^{|\mu\nu} \doteq \frac{\pd^2 f(\vec{x}, \vec{k})}{\pd k_\mu\pd k_\nu}.
\end{gather}
Since $\vec{k} = \del\theta(\vec{x})$, one has
\begin{gather}
f_{,\mu}\boldparen{\vec{x}, \vec{k}(\vec{x})} = f_{|\mu} + f^{|\nu} k_{\nu|\mu},\\
k_{\nu|\mu} = k_{\nu,\mu} = k_{\mu,\nu} = \theta_{,\mu\nu} = \theta_{,\nu\mu}.
\end{gather}

\subsection{Inner products}

The inner product for scalar waves on $M^n$ is
\begin{gather}
\braket{\Psi | \Phi} = \int \dd^nx \,\sqrt{\mdet{g}(\vec{x})}\,\Psi^*(\vec{x}) \Phi(\vec{x}),
\end{gather}
and for $m$-dimensional vector waves on $M^n$,
\begin{gather}
\braket{\Psi|\Phi}_m \doteq 
\int \dd^nx \,\sqrt{\mdet{g}(\vec{x})}\,\gamma_{ab}(\vec{x}) \Psi^{a*}(\vec{x}) \Phi^b(\vec{x}).
\end{gather}
In particular, in the $\vec{x}$ representation,
\begin{gather}
\Psi^\dag(\vec{x}) \doteq \braket{\Psi|\vec{x}} = (\Psi_1^*, \Psi_2^*,\, \ldots, \Psi_m^*).
\end{gather}
We also use the dot product
\begin{gather}
\Psi \cdot \Phi \doteq \Psi^\dag(\vec{x}) \Phi(\vec{x}) = \Psi_a^*(\vec{x}) \Phi^a(\vec{x}).
\end{gather}

The inner product on the transverse space $M^{n-1}_\perp$ is defined for scalar fields as
\begin{gather}
\braket{\psi | \phi}^\perp \doteq 
\int \dd^{n-1}\varrho\,\sqrt{\mdet{h}(\zeta, \vec{\varrho})}\,
\psi^*(\zeta, \vec{\varrho}) \phi(\zeta, \vec{\varrho})
\end{gather}
and for the $N$-dimensional ``active'' parts of the vector fields (\Sec{sec:act}) as
\begin{gather}
\braket{a | b}_N^\perp \doteq \int \dd^{n-1}\varrho\,\sqrt{\mdet{h}(\zeta, \vec{\varrho})}\,
a^*_s(\zeta, \vec{\varrho}) b^s(\zeta, \vec{\varrho}).
\end{gather}

\begin{widetext}
\section{Auxiliary calculations}
\label{app:aux}

Here we present some auxiliary calculations whose results are used in \Sec{sec:vector}. 

\subsection{Calculation of $\boldsymbol{\oper{K}}$ and $\boldsymbol{U}(\vec{x})$}
\label{app:aux1}

We start with \Eq{eq:K} for $\oper{K}$. This equation can be rewritten as follows:
\begin{align}
\oper{K} a
& = \frac{1}{2}\,\Xi^+(\oper{p}_{\mu}\mc{V}_H^{\mu} + \mc{V}_H^{\mu}\oper{p}_{\mu}- i\ell_\mu \mc{V}_H^\mu)\Xi a 
+ \new{\Xi^+ C_H\Xi}
\notag\\
& = -\frac{i}{2\mdet{g}^{1/4}}\,[\Xi^+(\mdet{g}^{1/4}\mc{V}_H^{\mu}\Xi a)_{,\mu}+\Xi^+ \mc{V}_H^{\mu}(\mdet{g}^{1/4}\Xi a)_{,\mu}] -\frac{i}{2}\,\Xi^+ \ell_\mu \mc{V}_H^\mu \Xi a 
+ \new{\Xi^+ C_H\Xi}
\notag\\
& =-\frac{i}{\mdet{g}^{1/4}}\,(\mdet{g}^{1/4})_{,\mu}\Xi^+\mc{V}_H^{\mu}\Xi  a-i \Xi^+\mc{V}_H^{\mu}\Xi  a_{,\mu}-\frac{i}{2}\,[\Xi^+(\mc{V}_H^{\mu}\Xi)_{,\mu}+\Xi^+\mc{V}_H^{\mu}\Xi_{,\mu}]a
 -\frac{i}{2}\,\Xi^+ \ell_\mu \mc{V}_H^\mu \Xi a
 + \new{\Xi^+ C_H\Xi}
 \notag\\
& =-\frac{i}{2}\,(\ln \sqrt{\mdet{g}})_{,\mu}V^{\mu} a-i V^{\mu}a_{,\mu} 
-\frac{i}{2}\,[V^{\mu}{}_{,\mu}-(\Xi^+)_{,\mu}\mc{V}_H^{\mu}\Xi + \Xi^+\mc{V}_H^{\mu}\Xi_{,\mu}
 +\Xi^+ \ell_\mu \mc{V}_H^\mu \Xi] a
 + \new{\Xi^+ C_H\Xi}
\notag\\
& =-i V^{\mu}a_{,\mu}-\frac{i}{2}\,V^{\mu}{}_{;\mu}a 
- \frac{i}{2}\,[\Xi^+\mc{V}_H^{\mu}\Xi_{,\mu}-(\Xi^+)_{,\mu}\mc{V}_H^{\mu}\Xi
 +\Xi^+ \ell_\mu \mc{V}_H^\mu \Xi] a + \new{\Xi^+ C_H\Xi},
\end{align}
where we used \Eq{eq:div} and introduced
\begin{gather}\label{eq:VVMU}
V^\mu \doteq \Xi^+ \mc{V}_H^{\mu} \Xi  
= (\Xi^+ \Deffr \Xi)^{|\mu} - \Xi^{+|\mu} \Deffr \Xi - \Xi^+ \Deffr \Xi^{|\mu} = \Lambda^{|\mu} - \Xi^{+|\mu} \Xi  \Lambda - \Lambda \Xi^+ \Xi^{|\mu} 
\approx \Lambda^{|\mu},
\end{gather}
where, at the end, we used $\Lambda\sim O(\epsilon)$ (\Sec{sec:actam}), so that the last two terms in \Eq{eq:VVMU} can be neglected.

From \Eq{eq:Xip}, we have $(\Xi^+)_{,\mu} = \delta^{-1} \Xi^{\sf H}_{,\mu} \vmetr + \Xi^{\sf +} \ell_\mu$. Using this and also \Eq{eq:vmugamma}, one obtains
\begin{gather}
\oper{K} a = -i V^{\mu}a_{,\mu}-\frac{i}{2}\,V^{\mu}{}_{;\mu}a - U a,
\quad
\new{U = U_0 - \Xi^+ C_H\Xi,
}
\\
\new{
U_0 = \frac{i}{2}\,\delta^{-1}(\Xi^{\sf H}\underline{\mc{V}}_H^{\mu}\Xi_{,\mu}-\Xi^{\sf H}_{,\mu}\underline{\mc{V}}_H^{\mu}\Xi)
= \delta^{-1}(\Xi^{\sf H}_{,\mu}\underline{\mc{V}}_H^{\mu}\Xi)_A,
}
\label{eq:U1}
\end{gather}
where the subscript $A$ denotes the anti-Hermitian part, as usual.

\subsection{Calculation of $\boldsymbol{U}_0(\vec{x}, \vec{k})$}
\label{app:aux2}

Let us also derive an alternative expression for $U_0$ [\Eq{eq:U1}] in terms of the partial derivatives $_{|\mu}$ instead of the ``full'' derivatives  $_{,\mu}$. Using \Eq{eq:comma}, one obtains $\Xi_{,\mu}=\Xi^{|\nu}k_{\nu,\mu}+\Xi_{|\mu}$. Then, \Eq{eq:U1} can be rewritten as follows:
\begin{gather}
U_0 = 
\frac{i}{2}\,\delta^{-1}(\Xi^{\sf H}\underline{\mc{V}}_H^{\mu}\Xi^{|\nu}-\Xi^{{\sf H}|\nu}\underline{\mc{V}}_H^{\mu}\Xi)
k_{\nu,\mu} +
\frac{i}{2}\,\delta^{-1}(\Xi^{\sf H}\underline{\mc{V}}_H^{\mu}\Xi_{|\mu}-\Xi^{\sf H}_{|\mu}\underline{\mc{V}}_H^{\mu}\Xi).
\end{gather}
By differentiating \Eq{eq:DXi} with respect to $k_{\mu}$, we obtain
\begin{gather}
\mc{V}_H^{\mu}\Xi 
=\Xi^{|\mu}\Lambda +\Xi \Lambda^{|\mu}-\Deffr\Xi^{|\mu}
\approx \Xi \Lambda^{|\mu}-\Deffr\Xi^{|\mu},
\end{gather}
where we used that $\Lambda$ (but not its partial derivatives) is small. Let us multiply this by $\vmetr$ to obtain
\begin{gather}\label{eq:Vsub}
\underline{\mc{V}}_H^{\mu}\Xi \approx \vmetr \Xi \Lambda^{|\mu}- \underline{\Deff}_H\Xi^{|\mu}, 
\quad
\Xi^{\sf H}\underline{\mc{V}}_H^{\mu} \approx \Lambda^{|\mu} \Xi^{\sf H} \vmetr - \Xi^{{\sf H}|\mu} \underline{\Deff}_H,
\end{gather}
where the latter equality is the complex conjugate of the former. Using these and $k_{\nu,\mu}=\theta_{,\nu \mu}=\theta_{,\mu \nu} = O(\epsilon)$, we further obtain, to the leading order,
\begin{align}
(\Xi^{\sf H}\underline{\mc{V}}_H^{\mu}\Xi^{|\nu}-\Xi^{{\sf H}|\nu}\underline{\mc{V}}_H^{\mu}\Xi)
k_{\nu,\mu}
& \approx(
\Lambda^{|\mu} \Xi^{\sf H} \vmetr \Xi^{|\nu} - \Xi^{{\sf H}|\mu} \underline{\Deff}_H \Xi^{|\nu}
-\Xi^{{\sf H}|\nu}\vmetr \Xi \Lambda^{|\mu} + \Xi^{{\sf H}|\nu}\underline{\Deff}_H\Xi^{|\mu}) k_{\nu,\mu} \notag \\
& =(
\Lambda^{|\mu} \Xi^{\sf H} \vmetr \Xi^{|\nu}-\Xi^{{\sf H}|\nu}\vmetr \Xi \Lambda^{|\mu}) k_{\nu,\mu} \notag \\
& =(\Lambda_{,\mu} \Xi^{\sf H} \vmetr \Xi^{|\mu} - \Xi^{{\sf H}|\mu}\vmetr\Xi\Lambda_{,\mu}) 
- (\Lambda_{|\mu}\Xi^{\sf H} \vmetr \Xi^{|\mu} - \Xi^{{\sf H}|\mu}\vmetr\Xi \Lambda_{|\mu})\notag\\
& \approx -(\Lambda_{|\mu} \Xi^{\sf H} \vmetr \Xi^{|\mu}-\Xi^{{\sf H}|\mu}\vmetr\Xi \Lambda_{|\mu}),
\end{align}
where we neglected terms of the second order in $\epsilon$. We also used the fact that ${\Lambda_s\boldparen{\vec{x},\vec{k}(\vec{x})}\lesssim O(\epsilon)}$ by definition of active modes and, in addition, $\Lambda_s$ are changing \textit{slowly}, so $\Lambda_{,\mu}\lesssim O(\epsilon^2)$. (Note that this is only true for $\Lambda_{,\mu}$ but not for $\Lambda_{|\mu}$.) Using \Eqs{eq:Vsub}, we also obtain similarly that
\begin{gather}
\Xi^{\sf H}\underline{\mc{V}}_H^{\mu}\Xi_{|\mu}-\Xi^{\sf H}_{|\mu}\underline{\mc{V}}_H^{\mu}\Xi
\approx \Lambda^{|\mu} \Xi^{\sf H} \vmetr \Xi_{|\mu} - \Xi^{{\sf H}|\mu} \underline{\Deff}_H \Xi_{|\mu}
-\Xi^{\sf H}_{|\mu}\vmetr \Xi \Lambda^{|\mu} + \Xi^{\sf H}_{|\mu}\underline{\Deff}_H\Xi^{|\mu}.
\end{gather}
Then,
\begin{multline}
U_0 = 
\delta^{-1}(
\Lambda_{|\mu} \Xi^{\sf H} \vmetr \Xi^{|\mu}-\Xi^{{\sf H}|\mu}\vmetr\Xi \Lambda_{|\mu}
-\Lambda^{|\mu} \Xi^{\sf H} \vmetr \Xi_{|\mu} + \Xi^{{\sf H}|\mu} \underline{\Deff}_H \Xi_{|\mu}
+\Xi^{\sf H}_{|\mu}\vmetr \Xi \Lambda^{|\mu} - \Xi^{\sf H}_{|\mu}\underline{\Deff}_H\Xi^{|\mu}
)/(2i) \\
= \delta^{-1}(
\Lambda_{|\mu} \Xi^{\sf H} \vmetr \Xi^{|\mu}
-\Lambda^{|\mu} \Xi^{\sf H} \vmetr \Xi_{|\mu}
+ \Xi^{{\sf H}|\mu} \underline{\Deff}_H \Xi_{|\mu}
)_A. \label{eq:aux51}
\end{multline}
As a reminder, the subscript $A$ denotes the anti-Hermitian part, and $\delta^{-1}$ is a unit matrix that only raises the mode index. Equation \eq{eq:aux51} is similar to the corresponding result derived in \Refs{my:qdirac, my:covar, phd:ruiz17}. Extra terms are included in those papers in the expressions for $U_0$, but they are of the second order in $\epsilon$ and could be neglected.

\subsection{Calculation of $\boldsymbol{\oper{\mc{G}}}$}
\label{app:aux3}

Consider \Eq{eq:mcG} for $\oper{\mc{G}}$ derived under the assumption of the quasioptical ordering. Here, we simplify that equation and derive the explicit formula \eq{eq:mcGvec}. First, note that
\begin{gather}
\Xi^+ \oper{\mcu{D}}_{H2}\Xi a
\approx \frac{1}{2}\,\Xi^+ \oper{p}_{\mu}\Theta_H^{\mu \nu}\oper{p}_{\nu}\Xi a
\approx -\frac{1}{2\mdet{h}^{1/4}}\,[\Xi^+ \Deffr^{|\sigma \sigmap}\Xi (\mdet{h}^{1/4}a)_{,\sigmap}]_{,\sigma},
\end{gather}
where we used \Eq{eq:DH2}. Also, using \Eq{eq:DH1} for $\oper{\mcu{D}}_{H1}$, one obtains
\begin{align}
\Xi^+ \oper{\mcu{D}}_{H1}\bar{\Xi}\bar{\Lambda}^{-1}\bar{\Xi}^+ \oper{\mcu{D}}_{H1}\Xi a
& \approx -(\Xi^+ \mc{V}_H^{\sigma}\bar{\Xi}\bar{\Lambda}^{-1}\bar{\Xi}^+ \mc{V}_H^{\sigmap}\Xi)\, a_{,\sigma \sigmap} \notag \\
& =-[\Xi^+ \mc{V}_H^{(\sigma}\bar{\Xi}\bar{\Lambda}^{-1}\bar{\Xi}^+ \mc{V}_H^{\sigmap)}\Xi]\, a_{,\sigma \sigmap}\notag \\
& \approx -\mdet{h}^{-1/4}[
\Xi^+ \mc{V}_H^{(\sigma}\bar{\Xi}\bar{\Lambda}^{-1}\bar{\Xi}^+ \mc{V}_H^{\sigmap)}\Xi(\mdet{h}^{1/4}a)_{,\sigmap}
]_{,\sigma}.
\end{align}
Here, in the first line, we substituted \Eq{eq:DH1} for $\oper{\mcu{D}}_{H1}$ and also \Eq{eq:sxp2} for $\oper{p}_\mu$. Then, we only kept the terms involving derivatives of $a$ along the perpendicular direction of the wave beam. The parenthesis in the indices denote symmetrization; namely, for any $A$ and $B$,
\begin{gather}\label{eq:symmetr}
A^{(\sigma}B^{\sigmap)} = \frac{1}{2}\,(A^{\sigma}B^{\sigmap} + A^{\sigmap}B^{\sigma}).
\end{gather}
The metric factor $\mdet{h}$ has been added to keep the operator self-adjoint under the inner product \eq{eq:innerperp2}. (Accounting for the inhomogeneity of $\boldsymbol{h}$ in $\oper{\mc{G}}$ is beyond the accuracy of our theory. However, keeping $\oper{\mc{G}}$ self-adjoint is convenient and physically meaningful, for it is known that the exact operator is not responsible for dissipation.) Hence,
\begin{gather}
\oper{\mc{G}}\approx -\frac{1}{2\mdet{h}^{1/4}}\,[\Phi^{\sigma \sigmap}(\mdet{h}^{1/4}a)_{,\sigmap}]_{,\sigma},
\quad
\Phi^{\sigma \sigmap} = \Xi^+ \Deffr^{|\sigma \sigmap}\Xi 
- 2\Xi^+ \Deffr^{(|\sigma}\bar{\Xi}\bar{\Lambda}^{-1}\bar{\Xi}^+ \Deffr^{|\sigmap)}\Xi.
\end{gather}
In order to simplify the expression for $\Phi^{\sigma \sigmap}$, note that
\begin{gather}
\bar{\Xi}^+ \Deffr^{|\sigmap}\Xi 
= (\bar{\Xi}^+ \Deffr\Xi)^{|\sigmap}
-\bar{\Xi}^{+|\sigmap}\Deffr\Xi
-\bar{\Xi}^+ \Deffr\Xi^{|\sigmap}
=(\bar{\Xi}^+ \Xi \Lambda)^{|\sigmap}
-\bar{\Xi}^{+|\sigmap}\Xi \Lambda 
-\bar{\Lambda}\bar{\Xi}^+ \Xi^{|\sigmap}
\approx \bar{\Lambda}\bar{\Xi}^{+|\sigmap}\Xi,
\end{gather}
where we used $\bar{\Xi}^+ \Xi = \mathbb{0}$ (and thus $\bar{\Xi}^+ \Xi^{|\sigmap} = -\bar{\Xi}^{+|\sigmap}\Xi$) and neglected terms proportional to $\Lambda = O(\epsilon_\lVert)$. Hence,
\begin{gather}
\Xi^+ \Deffr^{|\sigma}\bar{\Xi}\bar{\Lambda}^{-1}\bar{\Xi}^+ \Deffr^{|\sigmap}\Xi 
= \Xi^+ \bar{\Xi}^{|\sigma}\bar{\Lambda}\bar{\Lambda}^{-1}\bar{\Lambda}\bar{\Xi}^{+|\sigmap}\Xi 
= \Xi^+ \bar{\Xi}^{|\sigma}\bar{\Lambda}\bar{\Xi}^{+|\sigmap}\Xi.
\end{gather}
Using \Eqs{eq:Xiprop} for $\Xi$ and $\bar{\Xi}$ like we did above, \Eq{eq:aux1003} for $\Deffr$, and the fact that $\Lambda = O(\epsilon)$, we finally obtain
\begin{align}
\Phi^{\sigma \sigmap}
& = 
\Xi^+ (\Xi \Lambda \Xi^+ 
+ \bar{\Xi} \bar{\Lambda} \bar{\Xi}^+ )^{|\sigma \sigmap}\Xi 
- 2\Xi^+ \bar{\Xi}^{(|\sigma}\bar{\Lambda}\bar{\Xi}^{+|\sigmap)}\Xi 
\notag \\
& = 
\Xi^+ (\Xi^{|\sigma} \Lambda \Xi^+ 
+\Xi \Lambda^{|\sigma} \Xi^+ 
+\Xi \Lambda \Xi^{+|\sigma}+\bar{\Xi}^{|\sigma} \bar{\Lambda} \bar{\Xi}^+ 
+\bar{\Xi} \bar{\Lambda}^{|\sigma} \bar{\Xi}^+ 
+\bar{\Xi} \bar{\Lambda} \bar{\Xi}^{+|\sigma})^{|\sigmap}\Xi 
-2\Xi^+ \bar{\Xi}^{(|\sigma}\bar{\Lambda}\bar{\Xi}^{+|\sigmap)}\Xi 
\notag \\
& \approx 
 \Xi^+ (\Xi^{|\sigma} \Lambda^{|\sigmap} \Xi^+ 
 +\Xi^{|\sigmap} \Lambda^{|\sigma} \Xi^+ 
 +\Xi \Lambda^{|\sigma\sigmap} \Xi^+ 
 +\Xi \Lambda^{|\sigma} \Xi^{+|\sigmap}
 +\Xi \Lambda^{|\sigmap} \Xi^{+|\sigma}
 +\bar{\Xi}^{|\sigma} \bar{\Lambda} \bar{\Xi}^{+|\sigmap}
 +\bar{\Xi}^{|\sigmap} \bar{\Lambda} \bar{\Xi}^{+|\sigma})\Xi 
 -2\Xi^+ \bar{\Xi}^{(|\sigma}\bar{\Lambda}\bar{\Xi}^{+|\sigmap)}\Xi
 \notag \\
& =
 \Lambda^{|\sigma \sigmap}
 +\Xi^+ \Xi^{|\sigma} \Lambda^{|\sigmap}
 +\Xi^+ \Xi^{|\sigmap} \Lambda^{|\sigma}
 -\Lambda^{|\sigma} \Xi^+ \Xi^{|\sigmap}
 -\Lambda^{|\sigmap} \Xi^+ \Xi^{|\sigma}
 \notag \\
& \approx
 \Lambda^{|\sigma \sigmap} + [\Xi^+ \Xi^{(|\sigma}, V^{\sigmap)}],
\end{align}
where we substituted $\Lambda^{|\sigma} \approx V^{\sigma}$ [\Eq{eq:VVMU}]. The quasioptical approximation implies that $V^{\sigmap}$ is close to a scalar matrix, so the commutator $[\Xi^+ \Xi^{(|\sigma}, V^{\sigmap)}]$ can be neglected. Hence, $\Phi^{\sigma \sigmap} \approx \Lambda^{|\sigma \sigmap}$.

\mbox{}

\end{widetext}


\begin{thebibliography}{10}
\bibitem{foot:thereof}
It is not a purpose of this paper to provide a comprehensive list of existing
  approaches to reduced modeling of EM waves in plasmas, so the above reference
  list is not intended as exhaustive.
\bibitem{foot:raycon}
This does not extend to ray-tracing codes. For example, a ray-tracing code
  \textsc{raycon} \cite{book:tracy} captures mode conversion by using an
  approximate yet somewhat generic analytic solution that applies in the limit
  of a narrow region of resonant coupling. A related ray-tracing model with
  mode conversion was also proposed in \Refs{phd:ruiz17, my:covar}.
\bibitem{foot:mc}
\new{Mode conversion is understood here as the exchange of action between
  near-resonant but non-intersecting branches of the dispersion relation
  \cite{book:tracy}. Mathematically, this process is similar to quantum
  tunneling. Adiabatic transformations of a wave mode along continuous
  dispersion curves, such as the X--B transformation \cite{book:stix}, are not
  mode conversion in the true sense of the term and can also be simulated using
  already existing reduced models.}
\bibitem{foot:covar}
\new{This also facilitates the aforementioned applications to related problems
  in general relativity \cite{foot:marius}, as will be discussed in future
  publications.}
\bibitem{foot:orderone}
\new{Eventually (\Sec{sec:xgo}), we shall choose $\vec{k}(\vec{x})$ such that
  $D\boldparen{\vec{x}, \vec{k}(\vec{x})}$ be small. But that will characterize
  $\vec{k}(\vec{x})$ rather than the function $D$ itself, which is considered
  order-one when evaluated on general $(\vec{x}, \vec{p})$.}
\bibitem{foot:herm}
Strictly speaking, \Eq{eq:weq} can always be multiplied by a constant matrix to
  ensure that the new dispersion operator satisfies the ordering
  \eq{eq:ordering} at a given location. Hence, the only nontrivial requirement
  that we actually impose here is that such $\Deffi$ does not grow too much as
  the wave propagates.
\bibitem{foot:deff}
In principle, $\oper{\mc{D}}_A$ produces effects qualitatively different from
  those produced by $\oper{\mc{D}}_H$, so one might argue that the accuracy of
  its approximation does not necessarily need to equal that of
  $\oper{\mc{D}}_H$. Besides, $\Deffi$ is often more abrupt than $\Deffr$, so
  retaining $\smash{\mc{V}_A^\sigma}$ in \Eq{eq:Dpp} may in fact lead to a
  better model in principle. However, in such cases, the Weyl symbol $D$ of the
  original dispersion operator is typically known only to the zeroth order
  (\Sec{sec:statw}), and thus so is $\Deffi$. Hence, retaining first-order
  corrections the damping coefficient does not seem warranted.
\end{thebibliography}

\end{document}